\documentclass[12pt, preprint]{aastex}
\usepackage{amsmath}
\usepackage{hyperref}
 \shorttitle{Magnetized spherically symmetric accretion}
\shortauthors{Roman Shcherbakov}
\begin{document}
\title{SPHERICALLY SYMMETRIC ACCRETION FLOWS: MINIMAL MODEL WITH MHD TURBULENCE}
\author{Roman V. Shcherbakov}
\affil {Harvard-Smithsonian Center for Astrophysics, 60 Garden
Street, Cambridge, MA 02138} \email{rshcherbakov@cfa.harvard.edu, shcher@gmail.com}

\begin{abstract}
The first spherical accretion model was developed 55 years ago, but the theory is yet far from being complete. The real accretion
flow was found to be time-dependent and turbulent. This paper presents the minimal MHD spherical accretion model that separately
deals with turbulence. Treatment of turbulence is based on simulations of several regimes of collisional MHD. The effects of
freezing-in amplification, dissipation, dynamo action, isotropization, and constant magnetic helicity are self-consistently
included. The assumptions of equipartition and magnetic field isotropy are released. Correct dynamics of magnetized flow is
calculated. Diffusion, convection, and radiation are not accounted for. Two different types of Radiatively Inefficient accretion flows are found: a
transonic non-rotating flow (I), a flow with effective transport of angular momentum outward (II). Non-rotating flow has an
accretion rate several times smaller than Bondi rate, because turbulence inhibits accretion. Flow with angular momentum transport
has accretion rate about 10-100 times smaller than Bondi rate. The effects of highly helical turbulence, states of outer
magnetization, and different equations of state are discussed. The flows were found to be convectively stable on average, despite
gas entropy increases inward. The proposed model has a small number of free parameters and the following attractive property. 
Inner density in the non-rotating magnetized flow was found to be several times lower than density in a non-magnetized accretion. 
Still several times lower density is required to explain the observed low IR luminosity and low Faraday rotation measure of accretion onto Sgr A*.
\end{abstract}


\section{INTRODUCTION}
Dynamics of magnetized accretion flows is a major topic of astrophysical research. The problem can be solved with two different
approaches: numerical and analytical. Each of them has specific difficulties, so these methods can be applied together for a
better result.

Realistic numerical simulations require a lot of computational time to model even the isotropic case \citep{lazarian_turbulence}.
Convergence of properties of the isotropic turbulence is reached only when computational domain has more than $1024$ cells in
each dimension \citep{reynolds,biskamp03}. Non-isotropic simulations of this size were not performed. It is also very difficult
to model the system with large range of scales. The system then possesses vastly different timescales. Existing simulations of
accretion flows are either axisymmetric \citep{mckinney} or consider a rather small domain close to the object \citep{hawley,igumen06}. In addition, simulations
should be run for sufficiently long time or several runs should be made to obtain average quantities, e.g. accretion rate, power
of emitted radiation.

Analytical models do not suffer from a need to average, if they are based on averaged quantities. However, to build a reasonable
model is itself difficult. No unified method exists to combine insights in physics and mathematics into a perfect analytical
model. That is why the zoo of approximations of astrophysical flows is so huge.

 In particular, many analytical treatments were devised for accretion:
 spherically symmetric treatment \citep{bondi52,meszaros,coker,beskin}, standard disk
\citep{shakura}, Advection-Dominated Accretion Flow (ADAF) \citep{narayan} with its variation Hot Luminous Accretion Flow
\citep{yuan_hot}, Adiabatic Inflow-Outflow Solutions (ADIOS) \citep{blandford}, Convection Dominated Accretion Flow (CDAF)
\citep{narayan00,quataert_cdaf}, Jet-ADAF \citep{yuan_jet}. They are aimed to describe essentially the same process: axisymmetric
plasma inflow onto a compact source. Some models include the effects the others miss. Energy transport in CDAF, outflows in ADIOS
are the examples. Some effects are not treated properly in any approximation.

 Magnetic field is a main source of uncertainty and mistakes in theory of accretion flows. Two
assumptions are usually posed to incorporate it into the model. Firstly, magnetic field is considered to be isotropic
\citep{coker,narayan}. Then magnetic pressure and magnetic energy density may be put \citep{narayan} into the dynamical
equations. Secondly, the ratio of magnetic field energy density to gas thermal energy density is set to constant. This is called
thermal equipartition assumption. These two ideas are at least unproven or may even not work. Magnetic field is predominantly
radial in spherical inflow \citep{schwa} because of freezing-in condition and predominantly toroidal in disk \citep{hawley}
because of magnetorotational instability.

In a good model direction and strength of the magnetic field should be determined self-consistently. Non-isotropy of magnetic
field requires special dynamics. Dynamical equations were partially derived more than 20 years ago \citep{scharlemann}, but did
not receive much attention or were even considered erroneous \citep{beskin}.

Such a model may offer a natural explanation of certain accretion patterns. Accretion onto Sgr A* gives an excellent opportunity
for testing. Our Galaxy is proven to host a Supermassive Black Hole (SMBH) named Sgr A* in its center \citep{Ghez,shen}. This
black hole accretes matter and emits radiation with characteristic low-luminosity spectrum \citep{narayan98}. This spectrum was
satisfactory explained with the combination of two models: jet or non-thermal \citep{yuan} radio-emission and X-Rays with IR
radiation coming from conventional ADAF flow. However, the large number of free parameters allows one to fit any spectrum well. Model with no free parameters left is an ultimate goal of the ongoing study.

Partial progress in building a self-consistent accretion model is made in this paper, which is organized as follows. Averaged
spherical MHD model with turbulence is devised in Section~\ref{sec_model}. Approximate model employs the characteristic length scale about the size of the region of interest. Coefficients are taken from several hydrodynamic and MHD simulations. External sources sustain turbulence at large radii, whereas 
turbulence is self-sustained in the converging flow at small radii. Necessary boundary conditions are discussed in
Section~\ref{sec_boundary} for general flow and for Sgr A*. Results in Section~\ref{sec_results} are followed by the discussion
of the model in Section~\ref{sec_discussion}. Observational implications in Section~\ref{sec_observations} are supplemented with
prospects for future work and Conclusion in Section~\ref{sec_conclusion}. Paper has several appendices.

\section{SPHERICAL MODEL}\label{sec_model}

I base all calculations on Magneto Hydrodynamic system of equations \citep{landau8}. The viscous terms are retained where they do
not vanish in the limit of vanishing viscosity. The quantities in the following equations are fully dependent on time and
coordinates. General mass flux equation reads
\begin{equation}\label{gen_continuity}
\frac{\partial\rho}{\partial t}+ {\bf  \nabla}(\rho {\bf  V}) = 0,
\end{equation}
where $\bf  V$ is fluid velocity. Force balance is described by
Navier-Stokes equation
\begin{equation}\label{gen_Euler}
\frac{\partial {\bf V}}{\partial t}+ ({\bf  V~\nabla}){\bf V}=-\frac{{\bf \nabla}  p}{\rho}-{\bf \nabla}\phi_{\rm g}-\frac{[{\bf
B \times [\nabla\times B]}]}{ 4 \pi\rho}+\nu \triangle {\bf  V},
\end{equation}
where $\phi_{\rm g}$ is gravitational potential, $\nu$ is kinematic viscosity. The last term is responsible for finite energy
dissipation through Kolmogorov cascade \citep{landau6}. Momentum equation is a combination of equations (\ref{gen_continuity})
and (\ref{gen_Euler})
\begin{equation}\label{gen_momentum}
\frac{\partial (\rho V_i)}{\partial t}= -\frac{\partial}{\partial x_k}\left(p \delta_{i k}+\rho V_i
V_k+\frac{1}{4\pi}\left(\frac12 B^2 \delta_{i k}-B_i B_k\right)\right)-\frac{\partial\phi_{\rm g}}{\partial x_i} + \nu(\triangle
{\bf  V})_i.
\end{equation}
Energy equation
\begin{equation}\label{gen_energy}
\frac{\partial}{\partial t} \left(\frac{\rho {\bf V}^2}{2}+\rho\varepsilon+\frac{{\bf  B}^2}{8\pi}\right)= -{\bf
\nabla}\left(\rho {\bf  V}\left(\frac{{\bf  V}^2}{2}+\phi_{\rm g} + w \right)+\frac{1}{4\pi}[{\bf  B\times[ V \times B]}]+ \rm
\bf viscous \right)
\end{equation}
includes information about the equation of state. Here $\varepsilon$ is gas internal energy density, $w=\varepsilon+\int dp/\rho$
is gas specific enthalpy. Viscous term is responsible for diffusion. Magnetic field evolution is described by induction equation
\begin{equation}\label{gen_induction}
\frac{\partial {\bf B}}{\partial t}={\bf \nabla}\times[{\bf V}\times{\bf B}]+\nu_M \triangle {\bf B}
\end{equation}
with magnetic diffusivity $\nu_M.$ Magnetic field is solenoidal as well as incompressible random velocity field:
\begin{equation}\label{gen_solenoid}
{\bf \nabla B}=0, \qquad {\bf \nabla u}=0.
\end{equation}

\subsection{Dynamics}
Spherical accretion is the simplest pattern of all symmetric setups. We need to solve the basic model first to move then to a
more realistic pattern. Construction of the minimal maximally symmetric model is the subject of the following study.

I employ the natural for the problem spherical coordinates $(r, \theta, \phi)$ and average over angular variables $(\theta,
\phi).$ The results depend only on the radial variable $r$ and not on time $t$ in the assumption that angular averaging is the
same as time averaging. I need now to determine the essential quantities and derive the closed system of equations on them.

 Essential quantities of
a non-magnetized solution in \citet{bondi52} are the inflow speed $v(r),$ density $\rho(r),$ and temperature $T(r)$. Turbulent
magnetized case requires several more. As I release the assumption of isotropy, there are two special directions: along the
radial vector ${\bf  e}_r$ and perpendicular to the radial vector. To describe realistic Magneto Hydrodynamic turbulence, I need
at least $6$ quantities: squares of radial and perpendicular magnetic fields $B_{r}^2$ and $B_\perp^2$, squares of radial and
perpendicular random fluid speeds $u^2$ and $u_\perp^2$, characteristic length scale $L,$ and dimensionless magnetic helicity
$\xi.$ The last quantity will be described in detail in the corresponding subsection~\ref{subs_helicity}. For simplicity I
consider random velocity to be isotropic and denote it as $u(r)$.

 Total velocity of a fluid parcel
 \begin{equation}\label{decomposition}
{\bf  V}(r,\theta,\phi, t)= v(r) {\bf  e}_r+ {\bf  u}(r,
\theta,\phi, t)
\end{equation}
 is a sum of averaged inflow speed $v(r)$ and
instantaneous random velocity ${\bf u}(r, \theta, \phi, t),$ where by definition angular average of turbulent velocity vanishes
\begin{equation}\label{u_aver}\int {\bf  u}(r,\theta,\phi,t)d\Omega=0.
\end{equation} General continuity equation (\ref{gen_continuity}) can be averaged with
the aid of equations (\ref{gen_solenoid}) and (\ref{u_aver}) to
\begin{equation}\label{mdot_r}
4\pi\rho(r) v(r) r^2=\dot{M},
\end{equation}
where $\dot{M}$ is the mass accretion rate.

I derive the averaged force equation from general momentum equation (\ref{gen_momentum}). Tensor $\rho V_i V_k$ averages out into
the diagonal form $\rho v^2 \delta_{rr}+\rho u^2 \delta_{ik}/3.$ Because there are no sources of magnetic field (eq.
[\ref{gen_solenoid}]) and spherical geometry is assumed, no regular magnetic field exists. Following \citet{scharlemann}, I add
$B_r {\bf  \nabla B}/(4\pi\rho)$ to the radial magnetic force $F_r=[{\bf  B\times [\nabla\times B]}]_r/(4\pi\rho),$ average over
the solid angle, and then set $B_\phi=B_\perp$ and $B_\theta=B_\perp.$ Cross-terms with $(B_\theta B_r),$ $(B_\phi B_r),$ and
$(B_\phi B_\theta)$ cancel on average over the solid angle. Finally, I obtain
\begin{equation}\label{force_magnetic}
F_r=\frac{(r^4 B_{r}^2)'_r}{8\pi\rho r^4}-\frac{(r^2
B_\perp^2)'_r}{4\pi\rho r^2}
\end{equation} for the magnetic force. I denote by $()'_r$ radial derivatives. I omit bulk viscosity term that results from $\nu \triangle \bf
V.$ Paczynski-Wiita gravitational potential \citep{paczhynski}
\begin{equation}\phi_{\rm g}=-\frac{r_{\rm g} c^2}{2 (r-r_{\rm g})}
\end{equation} is used to imitate the effects of General
Relativity, where
\begin{equation}\label{r_g}
r_{\rm g}=\frac{2G M}{c^2} \end{equation} is a Schwarzschild radius of an object with mass $M$. I take gas pressure to be that of
an ideal gas $p=\rho R T/\mu,$ where $\mu$ is a mean molecular weight. Combining all the terms, I come to the averaged force
equation
\begin{equation}\label{Euler} v v'_r +\frac{r_{\rm g} c^2}{2 (r-r_{\rm g})^2} + \frac{R}{\mu}\frac{(\rho T)'_r}{\rho}
+ \frac{(\rho u^2)'_r}{3\rho}+\frac{(r^2 B_\perp^2)'_r}{4\pi\rho
r^2}-\frac{(r^4 B_{r}^2)'_r}{8\pi\rho r^4}=0.
\end{equation}

Averaged energy advection equation can be derived directly from general energy equation (\ref{gen_energy}). Enthalpy term should
include contribution from random fluid motions as well as from gas. Isotropic random motions of fluid exert isotropic pressure
$p_{\rm rand}=\rho u^2/3$ and have the internal energy density $\varepsilon_{\rm rand}=u^2/2.$ Total enthalpy $w$ is
\begin{equation}\label{enthalpy}
w=w_{\rm gas}+w_{\rm rand}, \quad {\rm where} \quad w_{\rm
gas}=\frac{R T(f_e a_e(T)+f_i a_i(T)+1)}{\mu}\quad {\rm and}\quad
w_{\rm rand}=\frac56u^2.
\end{equation}
Fractions of electrons $f_e\approx0.54$ and ions $f_i\approx0.46$ are calculated for a gas with twice solar abundance of
elements. Such high concentration of helium and metals was assumed by \citet{baganoff} for spectrum fitting of Sgr A*.
Correspondent mean molecular weight is $\mu\approx0.7{\rm g}~{\rm cm}^{-3}.$ Integral heat capacity per particle $a_e(T)$ and
$a_i(T)$ are different for electrons and ions. Ions are non-relativistic down to $r_{\rm g}$ \citep{narayan}. Therefore
$a_i(T)=3/2$. General expression \citep{chandra} should be used for thermal relativistic electrons $a_e(T)=\Theta^{-1}(
3K_3(\Theta^{-1})+K_1(\Theta^{-1}))/(4K_2(\Theta^{-1})-1).$ Here $\Theta=k T/m_e c^2$ is dimensionless temperature, $K_x(Y)$ are
modified Bessel functions of the second kind. Expression for non-relativistic enthalpy is
\begin{equation}\label{enthalpy_NR}
w_{NR}=\frac{5RT}{2\mu}+\frac56u^2.
\end{equation}
 It is valid in the limit $\Theta\ll 1.$ Time derivatives in energy equation (\ref{gen_energy}) vanish under averaging. Equation
takes the form  ${\bf \nabla q}=0,$ where ${\bf q}$ is the energy flux. Part of flux proportional to random velocity ${\bf u}$
averages out, because turbulence is incompressible and $u$ is zero on average (eq. [\ref{u_aver}]). Applying continuity relation
(\ref{mdot_r}), I finally obtain
\begin{equation}\label{advect} v v'_r +
\frac{r_{\rm g} c^2}{2 (r-r_{\rm g})^2}+ w'_r +
\frac{1}{2\pi}\left(\frac{B_\perp^2}{\rho}\right)'_r=0,
\end{equation} where again $B_\theta^2=B_\phi^2=B_\perp^2.$ I assumed the term $\int
[{\bf  B\times[u\times B]}]d\Omega$ to also be zero along with all
viscous energy transfer terms. I limit this study to Advection
Dominated flows by deliberately cutting off diffusion and
convection (see Appendix~\ref{section_convection}).

Subtracting force equation (\ref{Euler}) from energy advection
equation (\ref{advect}) I get the heat balance equation that reads
in non-relativistic limit
\begin{equation}\label{entropy} \frac{R}{\mu}\left(\frac32T'_r-\frac{\rho'_r}{\rho}T\right)+
\left(\left(\frac{u^2}{2}\right)'_r-\frac{\rho'_r}{\rho}\frac{u^2}{3}
\right) +\frac{\rho r^2}{4\pi} \left(\frac{B_\perp^2}{\rho^2
r^2}\right)'_r+\frac{1}{8\pi\rho r^4} (r^4 B_{r}^2)'_r=0,
\end{equation}
 similar to entropy conservation in hydrodynamics. Work done by gas is
represented by $-\rho'_r/\rho T.$ The first term has exactly the form of the second, if I make the substitution of the mean
square particles velocity
\begin{equation}\label{v_p}
v_p^2=\frac{3R T}{\mu}. \end{equation} Work done by the magnetic
field enters the expression as derivatives of $\rho$ and $r$ in
the magnetic part.

\subsection{Evolution of Turbulence}
Dynamics is the only part of ideal Bondi problem \citep{bondi52}. In reality, flow always has some small scale turbulence that
exerts back-reaction on the mean flow. The magnitude of back-reaction terms should be determined from additional equations that
describe the evolution of random magnetic field and fluid motions. Since no complete theory of turbulence exists, I make a lot of
approximations. The model is adjusted to agree with the results of several numerical simulations. I also apply analytical tests
similar to that in \citet{ogilvie} to assure the model reproduces the basic properties of observed turbulence.

I need non-ideal induction equation (\ref{gen_induction}) and Navier-Stokes equation (\ref{gen_Euler}) to derive how turbulence
evolves. My goal is to compound reasonable equations on average squares of radial magnetic field $B_{r}^2,$ perpendicular
magnetic field $B_\perp^2,$ isotropic velocity $u^2.$ I also need equations on characteristic length scale of turbulence $L$ and
dimensionless magnetic helicity $\xi.$

Radial part of induction equation (\ref{gen_induction}) easily gives the equation on $B_{r}^2$, when the former is multiplied by
$2B_{r}$ and averaged over the solid angle:
\begin{equation}\label{induction_inter}
2B_r \frac{\partial B_r}{\partial t}=2B_r[{\bf \nabla\times}[v{\bf e}_r{\bf \times B}]]_r+2B_r[{\bf \nabla\times[u\times
B]}]_r+2\nu_M B_r(\triangle{\bf  B})_r,
\end{equation}
where indices $()_r$ without primes denote the radial parts. The left-hand side vanishes as all time derivatives. The first term
on the right-hand side represents the uniform increase of magnetic field due to flux freezing. I combine it with the continuity
equation (\ref{mdot_r}) to eliminate $v$ derivatives. The second term is the dynamo action. It cannot be easily averaged.
Characteristic turbulence length scale $L$ may be used to approximate derivatives
\begin{equation}\label{averaging}
\frac{\partial B_i}{\partial x_k}\sim\frac{B_i}L {\bf  e}_k \quad {\rm and} \quad \frac{\partial u_i}{\partial
x_k}\sim\frac{u_i}L {\bf  e}_k, \quad {\bf e}_k {\rm -~unit~vector.}
\end{equation} Then we arrive at dynamo action
with characteristic timescale $\tau_{\rm dyn}=c_{Bu}\tau_{\rm edd}$ about eddy turn-over time $\tau_{\rm edd}=u/L.$ The averaged
expression is quadratic in magnetic field. I take coefficient to be $c_{Bu1}$ at any $B_i^2$ and $c_{Bu2}$ at any $B_i B_k$ with
$i\ne k.$ The final form of the dynamo term reads $2B_r[{\bf \nabla\times[u\times B]}]_r=(c_{Bu1}B_r^2 + c_{Bu2}
B_r(B_\theta+B_\phi))u/L,$ and characteristic
\begin{equation}\label{linear_field}B_r=\sqrt{B_r^2} \quad {\rm and} \quad
B_\theta=B_\phi=B_\perp=\sqrt{B_\perp^2} \end{equation} should be taken. The last term on the right-hand side of equation
(\ref{induction_inter}) represents magnetic field dissipation. Dissipation term $\nu_M \triangle {\bf B}$ of induction equation
(\ref{gen_induction}) is macroscopic in turbulence even for vanishing magnetic diffusivity $\nu_M$ \citep{biskamp03}. I
approximate radial dissipation to have a timescale $\tau_{\rm dyss}=c_{BB}\tau_{Ar}$ about Alfven timescale $\tau_{Ar}=v_{Ar}/L$.
The averaged expression is also quadratic in magnetic field. I take coefficient to be $c_{BB1}$ at any $B_i^2$ and $c_{BB2}$ at
any $B_i B_k$ with $i\ne k.$ Finally, $\nu_M B_r(\triangle {\bf  B})_r= v_A (c_{BB1} B_r^2+ c_{BB2}(B_\theta+B_\phi)B_r)/L.$
Collecting all the terms, I obtain
\begin{equation}\label{B_r}
\frac{v}{r^4}\frac{\partial(B_r^2 r^4)}{\partial
r}=\frac{-(c_{Bu1} B_r^2 + 2c_{Bu2} B_r B_\perp)u+(c_{BB1} B_r^2 +
2 c_{BB2} B_r B_\perp)v_{Ar}}{L}
\end{equation}
for the radial magnetic field in the absence of external energy sources.

Perpendicular part of induction equation (\ref{gen_induction}), for example $\theta$ part, gives the equation on $B_\theta^2$
when equation (\ref{gen_induction}) is multiplied by $B_\theta$ and averaged over the solid angle. The flux freezing condition
for perpendicular field is different from that for radial field: $B_\theta v r=\rm const$ represents perpendicular flux freezing.
I repeat the calculations made for radial field $B_r$ to find dynamo and dissipation terms. Dynamo term takes form
$(c_{Bu1}B_\theta^2 + c_{Bu2} B_\theta(B_\phi+B_r))u/L.$ Dissipation term is $v_{A\theta} (c_{BB1} B_\theta^2+
c_{BB2}(B_\phi+B_r)B_\theta)/L$ with perpendicular Alfven timescale for dissipation. Here I take $B_\perp^2=B_\theta^2=B_\theta
B_\phi=B_\phi^2.$ Finally, I obtain
\begin{equation}\label{B_perp_r}
v \rho^2 r^2\frac{\partial}{\partial
r}\left(\frac{B_\perp^2}{\rho^2
r^2}\right)=\frac{-((c_{Bu1}+c_{Bu2}) B_\perp^2 + c_{Bu2} B_\perp
B_r)u+((c_{BB1}+c_{BB2}) B_\perp^2 + c_{BB2}
 B_\perp B_r)v_{A\perp}}{L},
\end{equation} where continuity equation (\ref{mdot_r}) is used. Radial $v_{Ar}$ and perpendicular $v_{A\perp}$ Alfven
speeds and random velocity $u$ are
\begin{equation}\label{Alfven_speed}
v_{Ar}=\frac{\sqrt{B_r^2}}{\sqrt{4\pi\rho}}, \qquad v_{A\perp}=\frac{\sqrt{B_\perp^2}}{\sqrt{4\pi\rho}}, \qquad u=\sqrt{u^2}.
\end{equation} Coefficients $c_{Bu1},c_{Bu2},
c_{BB1}, c_{BB2}$ are yet to be determined.

Evolution equation for squared random fluid velocity $u^2$ can be found from momentum equation (\ref{gen_momentum}), when it is
multiplied by $2{\bf u}$ and averaged over the solid angle. Potential energy and pressure terms average out and only three terms
are left
\begin{equation}\label{u_intermediate}
2{\bf  u}\left(({\bf  V~\nabla}){\bf  V}+\frac{{\bf \nabla}(\rho
{\bf  V})}{\rho}\right)=2\frac{{\bf  u[B\times[\nabla\times
B]]}}{4\pi\rho}+2{\bf  u}\nu\triangle{\bf  u}.
\end{equation}
I apply the same averaging procedure as for magnetic field evolution equations (\ref{B_r}) and (\ref{B_perp_r}). The final result
is \begin{equation}\label{u_r} v \rho^{2/3}\frac{\partial}{\partial
r}\left(\frac{u^2}{\rho^{2/3}}\right)=\frac{c_{uu}u^3-(c_{uB1}v_A^2+(2c_{uB1}+c_{uB2})v_{A\perp}^2+2c_{uB2}(v_A
v_{A\perp}))u}{L},
\end{equation} with additional three coefficients $c_{uu}, c_{uB1}$
and $c_{uB2}.$ Some of these and other $c_{xx}$-like coefficients can be taken from numerical simulations of isotropic
turbulence, some of them can be inferred from analytical tests. They may not simply be set to convenient values like
\citet{ogilvie} did.

\subsection{Correspondence to Numerical Simulations}
Isotropic turbulence is studied quite thoroughly in numerical simulations. Some results are reproduced by a number of researchers
(see \citet{biskamp03} for the review). That is why we may believe in these results and base a model on them. Three simulations
of different turbulence regimes can provide four conditions that let us uniquely determine four combinations of coefficients
$c_{xx}.$ These regimes are decaying HD turbulence, decaying MHD turbulence, and dynamo growth of small seed magnetic field. I
assume then that $c_{xx}$ are constants independent of regime and extend the derived model to any anisotropic case.

Let me consider my model in isotropic incompressible case of box turbulence. In these settings $B_r^2=B_\theta^2=B_\phi^2.$
Squared magnetic field $B^2$ equals $B^2=3B_r^2.$ Transition to the co-moving frame of averaged inflow in turbulence evolution
equations (\ref{B_r}), (\ref{B_perp_r}), (\ref{u_r}) is done by stating $d/dt=-v\partial/\partial r.$ Now I should write time
derivatives instead of radius derivatives and set $r=\rm const,$ since matter is not moving anywhere from the box. I obtain
equations of evolution of isotropic turbulent Alfven speed $v_{A}$ and isotropic turbulent velocity $u$:
\begin{equation}\label{iso_turbulence}
\left(u^2\right)'_t=\frac{\hat{c}_{uB}v_{A}^2u-\hat{c}_{uu}u^3}{L},
\qquad \left(v_{A}^2\right)'_t=\frac{\hat{c}_{Bu}v_{A}^2
u-\hat{c}_{BB}v_{A}^3}{L}.
\end{equation} Here $v_A=\sqrt{B^2}/\sqrt{4\pi\rho}$ and $\rho=\rm const.$
Coefficients with hats are

\begin{alignat}{2}\label{hats}
\hat{c}_{Bu}&=c_{Bu1}+ 2c_{Bu2}, &\qquad
\hat{c}_{BB}&=\frac{c_{BB1}+2c_{BB2}}{\sqrt{3}},
\\\hat{c}_{uu}&=c_{uu1}, &\qquad
\hat{c}_{uB}&=c_{uB1}+c_{uB2}
 \nonumber
\end{alignat}
in terms of previously defined $c_{xx}.$

I have a freedom to set $L,$ because it enters the equations only in combinations $c_{xx}/L,$ but $c_{xx}$ are not yet
determined. For simplicity of further derivation I take $L(r)$ to be the effective size of energy containing eddies for isotropic
incompressible turbulence:
\begin{equation}\label{def_L}
u^2=\int^\infty_{2\pi/L}|u_k|^2 dk \quad {\rm and} \quad v_A^2=\int^\infty_{2\pi/L}|v_{Ak}|^2 dk.
\end{equation}

 Isotropic decay of hydrodynamic turbulence is the simplest simulation. The convenient constant of decay is Kolmogorov constant
$C_{HD}$. It is defined as
\begin{equation}\label{kolmogorov_hd}
C_{HD}=E_k k^{5/3}\epsilon^{-2/3} \quad {\rm with} \quad \epsilon=-\frac{d}{dt}\left(\frac{u^2}{2}\right) \quad {\rm and} \quad
E_k=\frac{|u_k|^2}{2},
\end{equation} where $E_k$ is energy spectrum, $\epsilon$ is a
decay rate. Kolmogorov constant was found to be
$C_{HD}\approx1.65$ in the large set of simulations \citep{sree}.
I substitute this number into equation (\ref{kolmogorov_hd}) and
evaluate the first integral in equation (\ref{def_L}) to find
\begin{equation}\label{cuu}
\hat{c}_{uu}=\frac{4\pi}{(3 C_{HD})^{3/2}}\approx 1.14
\end{equation} for isotropic equations (\ref{iso_turbulence}).

Isotropic decay of magneto hydrodynamic turbulence gives two conditions. MHD Kolmogorov constant is defined similarly to HD case
equation (\ref{kolmogorov_hd}) as
\begin{equation}\label{kolmogorov_mhd} C_{MHD}=E_k
k^{5/3}\epsilon^{-2/3} \quad {\rm with} \quad \epsilon=-\frac{d}{dt}\left(\frac{u^2+v_A^2}{2}\right) \quad {\rm and} \quad
E_k=\frac{|u_k|^2+|v_{Ak}|^2}{2}.
\end{equation} MHD turbulence is more difficult to model
numerically, but the value of $C_{MHD}\approx2.2$ is rather rigorous \citep{biskamp03}. In addition, kinetic energy was found to
decay in exactly the same rate as magnetic energy. Evaluation of the sum of two integrals (\ref{def_L}) with definitions
(\ref{kolmogorov_mhd}) and known $C_{MHD}$ yields
\begin{equation}\label{c_BB}
\hat{c}_{BB}-\hat{c}_{Bu}=\hat{c}_{uu}-\hat{c}_{uB}\approx
2\pi\left(\frac{2}{3C_{MHD}}\right)^{3/2} \approx1.05.
\end{equation}

Dynamo simulations explore the regime $v_A^2\ll u^2.$ Exponential growth of small magnetic field corresponds to some value of
coefficient $\hat{c}_{Bu}$ in equations (\ref{iso_turbulence}) as
\begin{equation}
B^2\propto \exp\left(\hat{c}_{Bu}\frac{u t}{L}\right).
\end{equation}
 External driving is purely mechanical for $v_A^2\ll u^2$, so external source
 of magnetic field does not alter the picture of field amplification by dynamo.
 Characteristic length scale in dynamo simulations is usually the size of energy  containing eddies $L$ consistent with definition (\ref{def_L}), so
  renormalization of length scale is not required. Older simulations \citep{kida} have found
$b=0.39$ that corresponds to $\hat{c}_{Bu}\approx 0.61.$ Later
results \citep{scheko} indicate a bit higher value
$\hat{c}_{Bu}\approx 0.7$ that I will use for my model. Finally,
\begin{alignat}{2}\label{coeff_hats}
\hat{c}_{Bu}&=0.70, &\qquad \hat{c}_{BB}&=1.75,
\\\hat{c}_{uu}&=1.14, &\qquad
\hat{c}_{uB}&=0.09.
 \nonumber
\end{alignat}

The values of four $\hat{c}_{xx}$ (eq. [\ref{coeff_hats}]) are not enough to obtain all seven coefficients $c_{xx}$ in equations
(\ref{B_r}), (\ref{B_perp_r}), (\ref{u_r}) with definitions (\ref{hats}). However, the application of common sense analytical
conditions to non-isotropic system of equations puts some additional constrains on $c_{xx}$ that allows me to complete the model
with as little guessing as possible.

Analytic tests are described in Appendix \ref{section_analytic_tests}. This completes the derivation and verification of
turbulence evolution equations (\ref{B_r}), (\ref{B_perp_r}), (\ref{u_r}) with coefficients
\begin{alignat}{4}\label{coefficients}
c_{BB1}&=3.03, &\qquad c_{BB2}&=0.00, &\qquad c_{Bu1}&=0.41,
&\qquad c_{Bu2}&=0.29,
\\c_{uu}&=1.14, &\qquad
c_{uB1}&=0.09, &\qquad c_{uB2}&=0.00\nonumber
\end{alignat}
that I obtain
summarizing equations (\ref{hats}), (\ref{coeff_hats}),
(\ref{c_Bu2}), and (\ref{c_uB2}). However, not all major effect
have been included so far.
\subsection{Magnetic Helicity}\label{subs_helicity}
Certain correlation called "magnetic helicity" may strongly influence magnetic field dissipation. This quantity is defined as
\begin{equation}\label{hel_def}
H=\int_V ({\bf  A~ B})dV,
\end{equation} where ${\bf A}$ is a vector potential with a
defined gauge condition \citep{biskamp00}. Time derivative of magnetic helicity is very small compared to the time derivative of
magnetic energy in high Reynolds number astrophysical plasma \citep{biskamp03}:
\begin{equation}\label{hel_proof}
\frac{dH}{dE_M}\frac{E_M}{H}\ll 1.
\end{equation}

Constancy of magnetic helicity defines the rules of selective decay. Magnetic energy $E_M$ decays in free turbulence down to
non-zero value, allowed by constant magnetic helicity $H=\rm const.$ The final force-free configuration has zero random kinetic
energy $E_K$ and has aligned current density and magnetic field ${\bf j}\upuparrows{\bf B}$ \citep{biskamp03}.

However, the derived system of turbulence evolution equations (\ref{incompressible_system}) and, therefore, equations
(\ref{B_r}), (\ref{B_perp_r}), (\ref{u_r}) cannot handle selective decay. Decay of magnetic energy must be modified in order to
have the transition to zero dissipation rate at certain $v_{Ar}$ and $v_{A\perp}$ as a function of magnetic helicity $H.$ First,
I should employ the proper magnetic helicity constancy. Then I should quantify the relation between critical $v_{Ar},$
$v_{A\perp},$ and $H.$

Let me consider the region $S$ that evolves together with the mean flow of fluid. This region has the constant angle boundaries
$\theta=\rm const$ and $\phi=\rm const$. Its radial elongation $L_{r}$ scales as inflow velocity: $L_{r}\propto v.$
 The region $S$ contains constant mass $m=\rm const$ of matter, because matter
flux through its boundaries is zero by definition. If I neglect diffusion by random velocity, frozen magnetic field lines do not
move through the boundaries of the region. Because of this, magnetic helicity in $S$ is constant $H=\rm const$ \citep{biskamp03}.

The simplest order of magnitude relation between magnetic energy
$E_M$ and $H$ is
\begin{equation}\label{hel_approx}
E_M L_H = H=\rm const
\end{equation}
in the region $S,$ where $L_H$ is magnetic helicity characteristic length scale \citep{biskamp03}. As magnetic field decays in
turbulence, $L_H$ grows according to equation (\ref{hel_approx}).

I can parametrize $L_H$ to be a fraction of $L:$
\begin{equation}\label{xi}
L_H=\xi L.
\end{equation}Volume of the region of interest $S$ is
\begin{equation}\label{volume}
V=\frac{m}{\rho}
\end{equation} with $m=\rm const.$ Total magnetic energy $E_M$ is
\begin{equation}\label{E_M}
E_M=\frac{V}{8\pi}(B_r^2+2B_\perp^2).
\end{equation}
I substitute relations (\ref{xi}), (\ref{volume}), and (\ref{E_M}) into equation (\ref{hel_approx}) and use the definitions
(\ref{Alfven_speed}) of Alfven velocities to come to
\begin{equation}\label{hel_r}
L(v_{Ar}^2+2v_{A\perp}^2)\xi=\rm const.
\end{equation}

Now I need to include $\xi$ into the turbulence evolution equations (\ref{B_r}), (\ref{B_perp_r}), (\ref{u_r}) so that they can
handle selective decay. The natural limit of $L_H$ growth is the characteristic size of energy containing eddies $L.$ So regime
$\xi\ll 1$ corresponds to non-helical turbulence and regime $\xi\sim 1$ to turbulence, where magnetic helicity significantly
inhibits dissipation. Regime $\xi\gg 1$ does not occur. The basic way to modify the equations is to decrease by a smooth
multiplier $f(\xi)<1$ magnetic field decay rate. For qualitative agreement with experiment \citep{biskamp03} I can employ
\begin{equation}\label{modifier}
f(\xi)=\exp(-\xi),
\end{equation} what means that magnetic energy dissipation
timescale becomes $\exp(\xi)$ times larger. Terms with both $u$ and one of $v_{Ar}$ and $v_{A\perp}$ in magnetic field evolution
equations (\ref{B_r}), (\ref{B_perp_r}) do not need to be modified, since random velocity energy decays to zero and these terms
do not matter. However, I multiply the term with both random velocity and Alfven speed in turbulent velocity evolution equation
(\ref{u_r}) by $\exp(-\xi)$ to make random velocity $u$ decay to zero.

\subsection{System of Equations with Source Terms}
With only minor corrections, the final system of equations can be written down. In general, turbulence has external sources of
energy that sustain finite magnetic and kinetic energies even in case of box turbulence. I can add source terms to incompressible
system (\ref{incompressible_system}) and consequently to the system of compressible equations (\ref{B_r}), (\ref{B_perp_r}),
(\ref{u_r}).

System (\ref{incompressible_system}) with coefficients (\ref{coeff_hats}) and (\ref{coefficients}), modifier (\ref{modifier}),
and source terms reads
\begin{mathletters}\label{incompressible_with_sources}
\begin{equation}\label{B_r_box_source}
\frac{d(v_{Ar}^2)}{d t}=\frac{(0.70 v_{Ar}^2 + 0.58
(v_{A\perp}-v_{Ar})v_{Ar})u-3.03 v_{Ar}^3\exp(-\xi)}{L} +
c_{0}\frac{v_p^3}{L},
\end{equation}
\begin{equation}\label{B_perp_box_source}
\frac{d(v_{A\perp}^2)}{d
t}=\frac{(0.70v_{A\perp}^2+0.29(v_{Ar}-v_{A\perp})v_{A\perp})u-3.03v_{A\perp}^3\exp(-\xi)}{L}
+ c_{1}\frac{v_p^3}{L},
\end{equation}
\begin{equation}\label{u_box_source}
\frac{d(u^2)}{d t}=\frac{0.09(v_{Ar}^2+2v_{A\perp}^2)u\exp(-\xi)
-1.14u^3}{L} + c_{2}\frac{v_p^3}{L},
\end{equation}
\end{mathletters}
where $v_p$ is the mean square particles speed (eq. [\ref{v_p}]) and $c_0,$ $c_1,$ and $c_2$ are dimensionless coefficients.
These coefficients determine the rates of external energy input into turbulent fields.

 I denote by $\sigma$ the ratio of total turbulent energy to thermal energy:
\begin{equation}\label{magnetization}
\sigma=\frac{E_K+E_M}{E_{\rm th}}, \quad{\rm so~ that}\quad\sigma\frac{3 R T}{2\mu}=\sigma
\frac{v_p^2}2=\frac{u^2}2+\frac{v_{Ar}^2}2+v_{A\perp}^2.
\end{equation} Unlike conventional plasma magnetization, magnetization $\sigma$ with definition
(\ref{magnetization}) includes the energy of random fluid motions.

In the dynamic equilibrium of constant $v_{Ar},$ $v_{A\perp},$ $u$ and known $\xi$ system (\ref{incompressible_with_sources})
gives three algebraic equations for ratios $v_{Ar}/v_p,$ $v_{A\perp}/v_p,$ and $u/v_p$ as functions of $c_0,$ $c_1,$ and $c_2.$
To estimate $c_0,$ $c_1,$ and $c_2$ I take stationary driven isotropic turbulence with kinetic energy $E_K$ equal to magnetic
energy $E_M.$ Isotropic turbulence of interest has $v_{Ar}=v_{A\perp}=u/\sqrt{3}.$  Such turbulence occurs far from the central
object, where outer magnetization is a constant $\sigma_\infty.$ Solving system (\ref{incompressible_with_sources}) I obtain
using equation (\ref{magnetization})
\begin{equation}\label{coeff_sources}
c_0=c_1\approx0.124\sigma_\infty^{3/2}, \qquad
c_2=3c_0\approx0.371\sigma_\infty^{3/2}
\end{equation} in case $\xi=0$. I apply these values even to turbulence with
$\xi>0.$ Total external energy input $Q_{+}$ into $E_K$ and $E_M$ is \begin{equation}\label{energy_input} Q_{+}\approx0.742
\sigma_\infty^{3/2}\frac{v_p^3}{L}.
\end{equation} This energy adds up to thermal gas energy after being processed through turbulence.
However, I do not adjust my dynamical equations (\ref{Euler}) and (\ref{advect}) for $Q_{+}.$ I self-consistently omit external heating
and radiative or diffusive cooling. This omission is physically justified sufficiently far from the central object, where cooling
$Q_-$ balances external heating $Q_+$. It is also justified in the inner region, where both $Q_+$ and $Q_-$ are negligible compared to the internal driving and energy advection. Internal driving represents build-up of self-sustained turbulence in a converging flow due to conservation of magnetic flux \citep{coker}.

Only the size $L$ of energy containing eddies should be specified to complete the derivation of closed system of equations. In
the case when energy input $Q_+$ does not matter, the problem has only one relevant scale that is the size of the system $r.$
Therefore, I can set $L$ to be the fraction of radius
\begin{equation}\label{L_rxx}
L=\gamma r
\end{equation} with the proportionality constant $\gamma$ about
unity. However, energy input from external sources $Q_+$ is relatively large far from the central source. This causes medium with
constant $Q_+,$ constant $v_p,$ and constant $\sigma_\infty$ to have constant size of largest eddies
\begin{equation}\label{L_ryy}
L=L_\infty=\rm const
\end{equation} because of equation (\ref{energy_input}). This equality holds for radii larger than some
$r_0\approx L_\infty/\gamma.$ I introduce a function with a smooth
transition from relation (\ref{L_rxx}) for $r\ll r_0$ to relation
(\ref{L_ryy}) for $r\gg r_0$:
\begin{equation}\label{L_r}
L(r) = L_\infty\left(1-\exp\left(-\frac{\gamma
r}{L_\infty}\right)\right)
\end{equation}

\newcommand{\fullref}{(\ref{mdot_r}), (\ref{Euler}), (\ref{advect}),
(\ref{B_r}), (\ref{B_perp_r}), (\ref{u_r}), (\ref{hel_r}), (\ref{L_r})} This completes derivation and verification of 8 equations
\fullref with coefficients (\ref{coeff_hats}), (\ref{coefficients}), and (\ref{coeff_sources}) on 8 quantities $L(r),$ $\xi(r),$
$v(r),$ $u(r),$ $v_{Ar}(r),$ $v_{A\perp}(r),$ $T(r),$ $\rho(r)$ that are the characteristic turbulent length scale, normalized
magnetic helicity, matter inflow velocity, turbulent velocity, radial Alfven speed, perpendicular Alfven speed, temperature, and
density. I rewrite the equations once again in terms of named quantities:
\begin{mathletters}\label{system}
\begin{equation}\label{mdot_r1}
4\pi\rho  v r^2=\dot{M},
\end{equation}
\begin{equation}\label{Euler1} v v'_r +\frac{r_{\rm g} c^2}{2
(r-r_{\rm g})^2} + \frac{R}{\mu}\frac{(\rho T)'_r}{\rho} +
\frac{(\rho u^2)'_r}{3\rho}+\frac{(r^2 \rho v_{A\perp}^2)'_r}{\rho
r^2}-\frac{(r^4 \rho v_{Ar}^2)'_r}{2\rho r^4}=0,
\end{equation}
\begin{equation}\label{advect1} v v'_r +
\frac{r_{\rm g} c^2}{2 (r-r_{\rm g})^2}+ w'_r +\frac53u u'_r+ 2(v_{A\perp}^2)'_r=0 \quad {\rm with}
\end{equation}
$$
w=w_R=\frac{R T}{\mu}\left(0.54\frac{ 3K_3(\Theta^{-1})+K_1(\Theta^{-1})}{\Theta(4K_2(\Theta^{-1})-1)}+1.69\right)+\frac56u^2
\quad {\rm or}\quad w=w_{NR}=\frac{5RT}{2\mu}+\frac56u^2,$$

\begin{equation}\label{B_r1}
v\frac{(\rho v_{Ar}^2 r^4)'_r}{\rho r^4}=\frac{3.03
v_{Ar}^3\exp(-\xi)-(0.70 v_{Ar}^2 + 0.58
(v_{A\perp}-v_{Ar})v_{Ar})u}{L}-\frac{0.64}{L_\infty}\left(\frac{R
T_\infty \sigma_\infty}{\mu}\right)^{3/2},
\end{equation}
\begin{equation}\label{B_perp_r1}
v \rho r^2\left(\frac{v_{A\perp}^2}{\rho r^2}\right)'_r=\frac{3.03
v_{A\perp}^3\exp(-\xi)-(0.70 v_{A\perp}^2 +
0.29(v_{Ar}-v_{A\perp})v_{A\perp})u}{L}-\frac{0.64}{L_\infty}\left(\frac{R
T_\infty \sigma_\infty}{\mu}\right)^{3/2},
\end{equation}
\begin{equation}\label{u_r1}
v\rho^{2/3}\left(\frac{u^2}{\rho^{2/3}}\right)'_r=
\frac{1.14u^3-0.09(v_{Ar}^2+2v_{A\perp}^2)u\exp(-\xi)}{L}-\frac{1.93}{L_\infty}\left(\frac{R
T_\infty \sigma_\infty}{\mu}\right)^{3/2},
\end{equation}
\begin{equation}\label{hel_r1}
L(v_{Ar}^2+2v_{A\perp}^2)\xi=3L_\infty\xi_\infty\frac{R T_\infty
\sigma_\infty}{\mu},
\end{equation}
\begin{equation}\label{L_r1}
L =L_\infty\left(1-\exp\left(-\frac{\gamma
r}{L_\infty}\right)\right).
\end{equation}
\end{mathletters}Here $\Theta=k T/m_e c^2.$
Since my prescription for external driving of turbulence is $Q_+=\rm const,$ I take $v_p$ and $L$ to be constant in the source
terms. Relativistic $w_R$ (eq. [\ref{enthalpy}]) and non-relativistic $w_{NR}$ (eq. [\ref{enthalpy_NR}]) values of enthalpy $w$
are employed. In the next section I describe the values of boundary conditions and parameters for the equations I solve.

\section{BOUNDARY CONDITIONS AND PARAMETERS}\label{sec_boundary}
The system (\ref{system}) consists of 5 differential and 3 algebraic equations and should be integrated inward from some outer
boundary at $r_x$. This requires knowledge of at least eight constants. Seven of them are the values "at infinity" $L_\infty,$
$T_\infty,$ $\rho_\infty,$ $\xi_\infty,$ $u_\infty,$ $v_{Ar\infty},$ $v_{A\perp\infty}.$ The eighth is the accretion rate
$\dot{M}.$ It is usually determined by some extra condition and is not adjustable. I assume isotropic turbulence with $E_K=E_M$
at the outer boundary. Therefore,
\begin{equation}\label{velocity_inf}
v_{Ar\infty}=v_{A\perp\infty}=\left(\frac{R
T_\infty\sigma_\infty}{\mu}\right)^{1/2} \quad {\rm and}\quad
u_\infty=\left(\frac{3R T_\infty\sigma_\infty}{\mu}\right)^{1/2},
\end{equation} and I have one model parameter $\sigma_\infty$ instead of 3 velocities
$v_{Ar\infty},$ $v_{A\perp\infty},$ and $u_\infty$. Another adjustable parameter of the model is $\gamma$ that determines the
size of energy containing eddies $L$ near the object (eq. [\ref{L_r1}]).

Parameter $\gamma$ is not free in principle, but its value cannot be determined within the proposed theory. Neither there exist
anisotropic MHD simulations that could provide $\gamma.$ All simulations to date show $\gamma$ to be within $0.2\div2$
\citep{tennekes,landau6,biskamp03} in both HD and MHD case. I assume the same range of $\gamma$ in my calculations.

\subsection{Outer Medium Transition}
Bondi radius
\begin{equation}\label{bondi}r_B=r_{\rm g} \frac{c^2}{c_\infty^2}
\quad{\rm with}\quad c_\infty=\left(\frac{5 R
T_\infty}{3\mu}\right)^{1/2}
\end{equation} is  the natural length scale of the
spherical accretion flow \citep{bondi52}. Density $\rho$ and temperature $T$ of plasma are constant for radii $r\gg r_B,$ because
gravitational energy and gas regular kinetic energy are negligible there compared to gas internal energy \citep{bondi52}.
Averaged magnetic field and averaged random velocity are also constant for $r \gg r_B$, because constant external energy input
balances dissipation in this region. As a consequence, $\xi=\xi_\infty$ and $L=L_\infty$ for $r\gg r_B.$

I set the outer boundary at $r_x=3r_B,$ where matter is almost uniform. Length scale $L_\infty$ should be determined from known
external energy input $Q_+$ and outer magnetization $\sigma_\infty.$ However, $Q_+$ is not known. I assume for simplicity
$L_\infty=\gamma r_B,$ so that $L$ changes its behavior near $r_B$ together with temperature and density.

Bondi radius is about $r_B\approx3\times10^5r_{\rm g}$ for our Galactic Center \citep{Ghez}. The properties of gas at $3r_B$ are
somewhat constrained from observations. I take the values for uniformly emitting gas model with temperature
$T_\infty\approx1.5\times10^7$~K, electron and total number densities $n_{e\infty}=26\rm cm^{-3}$, $n_\infty=48\rm cm^{-3}$
\citep{baganoff} at $r_x=3r_B$ that corresponds to $5''$ in the sky. The presence of dense cold component can make the average
temperature much lower and the average density much higher \citep{cuadra}, but I am leaving these uncertainties for future
research.

Expanding and colliding hyperalfvenic stellar winds provide magnetic field into the region. Its strength near Bondi radius is not
known. Only the very general estimate can be made. Matter magnetization is likely to be lower than the saturation value of
$\sigma_\infty=1$. I take the values in the range $\sigma_\infty=0.001\div1$ to cover all reasonable magnetization states of
matter at $3r_B$. If magnetic field is rather a product of decay than dynamo amplification, then the local dimensionless helicity
$\xi$ may be close to unity. I cover the range $\xi_\infty=0.001\div0.5$ in simulations to determine the possible dynamical
significance of non-zero magnetic helicity.

\subsection{Transition to Rotationally Supported Flow}
The system of equations (\ref{system}) has the same property as spherically symmetric system of hydrodynamic equations
\citep{bondi52}: subsonic solution exists for all accretion rates $\dot{M}$ up to maximum $\dot{M}^*$, transonic
solution is valid for the only value $\dot{M}^*,$ and no solution exists for $\dot{M}>\dot{M}^*$. The solution with
\begin{equation}\label{cond_maxrate}
\dot{M}=\dot{M}^* {\rm (for~transonic~solution)} \end{equation} is preferable, because it has the highest rate of
energy transfer towards the equilibrium state of the system matter-SMBH. The same argument is valid for a general
hydrodynamic nozzle \citep{landau6}. It is reasonable to expect that maximum mass flux solution for system with
magnetic field (\ref{system}) also obeys the condition (\ref{cond_maxrate}). However, even small amount of
angular momentum can change the picture.

Every real astrophysical accretion flow has non-zero specific angular momentum at the outer boundary
\begin{equation}\label{l}
l=\lambda r_{\rm g} c, \quad{\rm or~equivalently,}\quad l=v_{K \rm
cir} r_{\rm cir},
\end{equation} where $r_{\rm cir}$ is a radius where matter
becomes rotationally supported and $v_{K \rm cir}$ is Keplerian velocity at $r_{\rm cir}.$ General Newtonian expression for
Keplerian velocity at radius $r$ is
\begin{equation}\label{kepler}
v_{K}=c \sqrt{\frac{r_{\rm g}}{2 r}}.
\end{equation} At larger radii $r>r_{\rm cir}$
angular momentum exerts relatively small force $F_l\propto l^2/r^3$ on plasma, since $F_l$ decreases with radius faster than
gravitational force $F_{\rm g}\propto r_{\rm g} c/r^2.$ Numerical simulations \citep{cuadra} suggest $r_{\rm cir}\sim
3\times10^3r_{\rm g}$ for our Galactic Center.

When angular momentum (eq. [\ref{l}]) is large, $\lambda\gg1,$ it should be able to travel outward through the outer
quasi-spherical solution by means of $r\phi$ component of stress tensor $t_{\alpha\beta}.$ The angular averaged form of
this component is
\begin{mathletters}
\begin{equation}\label{trphixx} t_{r\phi}= \frac{<B_r
B_\perp>_\Omega}{4\pi},
\end{equation} where I neglect the kinetic part for the estimate. It can be transformed with the aid
of Schwartz formula $<xy>\le\sqrt{<x^2>}\sqrt{<y^2>}$ into
inequality
\begin{equation}\label{trphi}
 t_{r\phi}\le \frac{B_r B_\perp}{4\pi}
\end{equation} with definitions (\ref{linear_field}) of rms $B_r$ and $B_\perp.$
\end{mathletters}

Let us take a disk \citep{shakura} with height $H$ and write the
angular momentum transfer equation as
\begin{mathletters}
\begin{equation}
\frac{d(r^2 H t_{r\phi})}{dr}=0.
\end{equation}
The result of integration is \citep{popham}
\begin{equation}\label{momentum} \dot{M}l=4\pi H r^2
t_{r\phi},\end{equation}
\end{mathletters}
in case of large dimensionless angular momentum $\lambda\gg1$
\citep{popham}. I take specific angular momentum $l$ from equation
(\ref{l}) and the accretion rate to be
\begin{equation}\label{mdot_disk}
\dot{M}=2\pi r H\rho v.
\end{equation}
I substitute angular momentum $l$ from relation (\ref{l}),
accretion rate $\dot{M}$ from equation (\ref{mdot_disk}), Alfven
speeds from definitions (\ref{Alfven_speed}), Keplerian velocity
from equation (\ref{kepler}), and inequality (\ref{trphi}) on
$t_{r\phi}$ into angular momentum transfer equation
(\ref{momentum}) to obtain
\begin{mathletters}\label{condition}
\begin{equation}\label{condition_a} \frac{v v_K}{v_A v_{A\perp}}\sqrt{\frac{r_{\rm
circ}}{r}}=2\chi, \qquad \chi\le1
\end{equation}
that should be valid at any radius $r.$ Sometimes, this inequality is valid for $r>r_{\rm cir}$ if it is valid at $r_{\rm cir},$
so that condition (\ref{condition_a}) can in some cases be simplified to
\begin{equation}\label{condition_b}
 \frac{v v_K}{v_A v_{A\perp}}\le2 \quad{\rm at}\quad r_{\rm cir}.
\end{equation}
\end{mathletters}
Height of the disk $H$ cancels out of final expression, thus conditions (\ref{condition}) are approximately valid even
for flows with $H\approx r.$ Such flows are likely to describe the realistic transition region from outer
quasi-spherical inflow to inner rotational solution. There are no extra degrees of freedom to put conditions on the
surface of compact object, so I consider an object to be effectively a black hole.

Condition of angular momentum transport (\ref{condition}) may be stronger than maximum accretion rate condition
(\ref{cond_maxrate}). This depends on the value of specific angular momentum $l$ and viscous $\alpha$ parameter
\citep{shakura}.  Viscous $\alpha$ is approximately $\alpha\sim\chi~\sigma$ according to my definitions
(\ref{magnetization}) and (\ref{condition_a}). If $\alpha\gtrsim0.5,$ then accretion proceeds without direct dynamical
effect of rotation \citep{narayan_adaf}. Thus, two types of solutions are possible:
\begin{itemize}
\item{maximum accretion rate solutions that describe radial flows with small angular momentum $l \lesssim c r_{\rm g}$
or large viscosity $\chi~\sigma\gtrsim0.5$ (subsection~\ref{maximum_rate_solution}),}
 \item{flows with the rotational support that work for large angular momentum
 $l \gg c r_{\rm g}$ and small viscosity $\chi~\sigma\lesssim0.5$ (subsection~\ref{physical_solution}).}
\end{itemize} The condition (\ref{condition}) gives a crude estimate of the inflow velocity and accretion rate $\dot{M},$ since it assumes specific angular
momentum to be constant down to $r_{\rm cir}.$ As matter travels to $r_{\rm cir},$ the amount of specific angular
momentum left becomes smaller. Nevertheless, I calculate the solutions with effective angular momentum transport using
condition (\ref{condition}) to illustrate the dependence of accretion rate on model parameters for the rotating flow.

\section{RESULTS}\label{sec_results}
\subsection{Maximum Rate Solution}\label{maximum_rate_solution}
Let me first disregard the angular momentum transport condition (\ref{condition}) and calculate the flow with small angular
momentum $l\ll r_{\rm g} c,$ when mean rotation is not dynamically important.

The system of equations I solve (\ref{system}) can be rewritten as
\begin{equation}
\frac{(F_i)'_r}{F_i}=\frac{N_i({\bf  F},r)}{D} \quad {\rm for}
\quad i=1..8.
\end{equation}
Here $F_i(r)$ are 8 functions I solve for, $N_i({\bf  F},r)$ are
function- and radius- dependent numerators, and
\begin{equation}
D=1-\frac{v^2}{V_s^2}
\end{equation} is a common denominator. Critical velocity $V_s$ is
\begin{equation}\label{sonic}
V_s^2=c_{sg}^2+2v_{A\perp}^2 \quad{\rm with}\quad c_{sg}^2=c_s^2+\frac{5u^2}{3}.
\end{equation}Effective sound speed $c_{sg}$ is equal to that of plasma with effective particles velocity
$v_{pg}^2=v_p^2+u^2.$

According to the maximum-rate condition (\ref{cond_maxrate}) I search for a smooth solution that has a sonic point at some radius
$r_s$. The condition at $r_s$ is $D(r_s)=0.$ Zero denominator requires all the numerators $N_i({\bf  F}, r)$ to be zero at $r_s.$
It can be shown from system (\ref{system}) that all eight conditions $N_i({\bf F}(r_s),r_s)=0$ collapse into just one, what
indicates that maximum accretion rate solution is smooth. Two equalities
\begin{equation}\label{maxrate}
D(r_s)=0 \quad{\rm and}\quad N_1({\bf  F}(r_s),r_s)=0
\end{equation}
give the missing 8-th condition on $\dot{M}$ for system (\ref{system}) and the sonic radius $r_s.$ Thus, I have 7 conditions at
the boundary at $3r_B$ and 1 condition somewhere in the region. I employ the shooting method to search for $\dot{M}$ and $r_s$
that satisfy the relation (\ref{maxrate}).

\begin{figure}[h]
\plotone{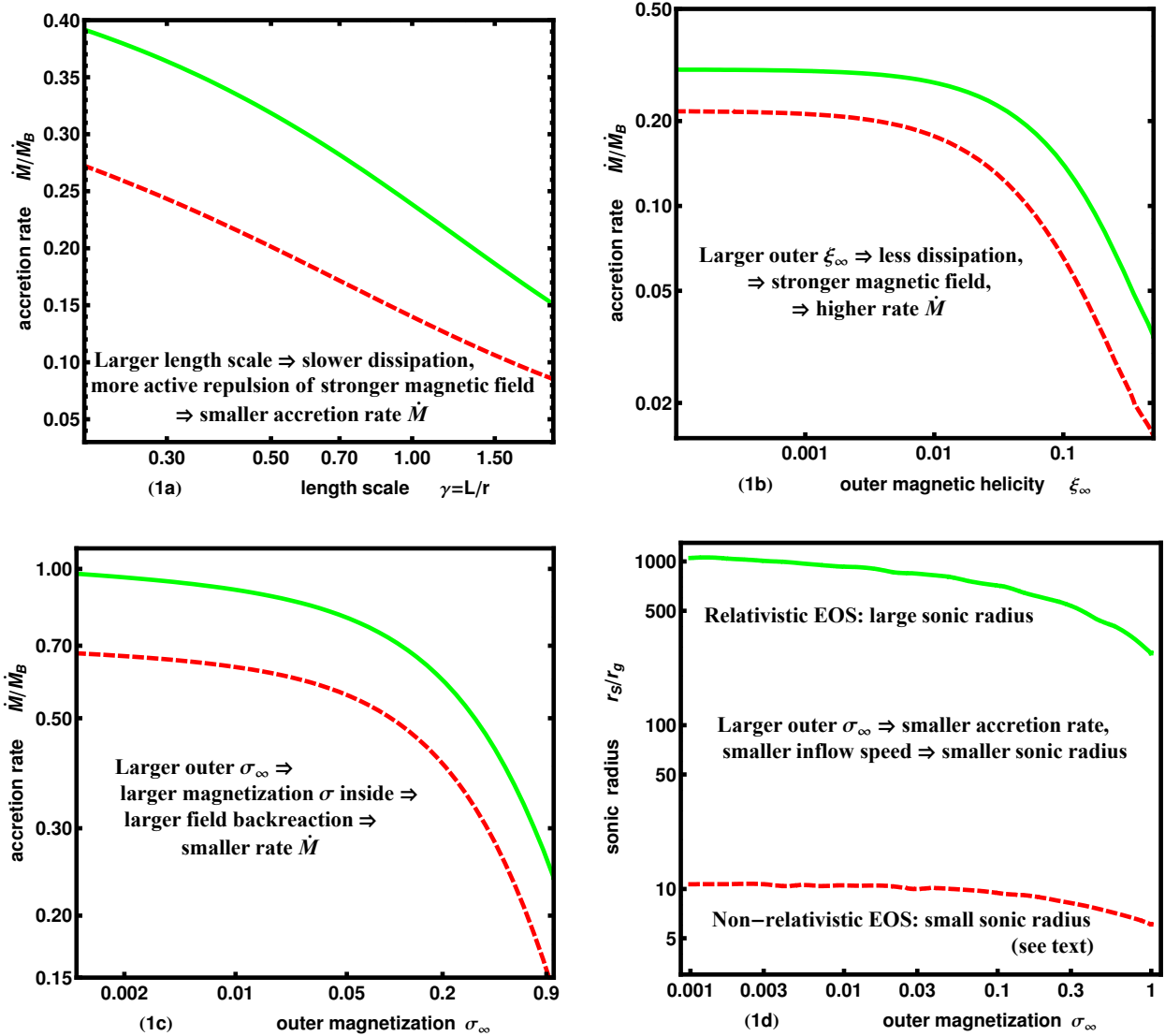}
 \caption{\textbf{Maximum accretion rate solution.} Dependence of the accretion rate in units of Bondi rate on dimensionless parameters:
characteristic length scale $\gamma$ (Fig.~\ref{fig2}a), outer magnetic helicity $\xi_\infty$ (Fig.~\ref{fig2}b), outer matter
magnetization $\sigma_\infty$ (Fig.~\ref{fig2}c). Dependence (Fig.~\ref{fig2}d) of sonic radius on outer magnetization
$\sigma_\infty$. I take the reference model to have the following values of parameters: $\gamma=1$, $\sigma_\infty=1$,
$\xi_\infty=0.025$. One parameter is varied to make one plot. Non-relativistic 1-T equation of state (dashed) verses relativistic
1-T equation of state (solid).}
 \label{fig2}
\end{figure}

I obtain the Bondi hydrodynamic model \citep{bondi52}, if I set all Alfven velocities and turbulent velocity to zero and use
non-relativistic prescription for enthalpy $w_{NR}$ (eq. [\ref{advect1}]). Therefore, the accretion rate $\dot{M}$ equals Bondi
accretion rate of monatomic gas
\begin{equation}\label{bondi_rate}
\dot{M}_B=\frac{\pi}{4}r_{\rm g}^2 c^4\rho_\infty
\left(\frac{3\mu}{5R
T_\infty}\right)^{3/2}\approx4\times10^{-6}M_\odot {\rm year}^{-1}
\end{equation} in the limiting case of no turbulence. The number
is calculated for the Black Hole in our Galactic Center with $r_{\rm g}=1.1\times10^{12}{\rm cm}$ \citep{Ghez},
$T=1.5\times10^7{\rm K},$ and $n\approx 48{\rm cm}^{-3}$ \citep{baganoff}. Accretion rate $\dot{M}$ appears to be lower than
$\dot{M}_B$ when turbulent energy is non-zero (Fig.~\ref{fig2}).

Inhibition of accretion by turbulence has the following explanation. First, energy of magnetic field increases inward, therefore
it exerts back-reaction force stopping matter \citep{schwa}. Second, magnetic field serves a very effective mechanism of energy
conversion from gravitational to thermal via dissipation of turbulence \citep{igumen02}. Larger thermal energy corresponds to
larger gas pressure that also stops matter. Within the deduced model I can estimate the actual decrease of accretion rate
$\dot{M}$ from Bondi value $\dot{M}_B.$

I take my reference model to have the values $\gamma=1,$ $\sigma_\infty=1,$ $\xi_\infty=0.025$ of, correspondingly, dimensionless
scale of turbulence, outer magnetization, and outer magnetic helicity. The found accretion rates are $0.14\dot{M}_B$ for
non-relativistic equation of state and $0.24\dot{M}_B$ for relativistic equation of state. I can now consider the whole ranges of
all three parameters and explain the observed correlations between them and accretion rate $\dot{M}$.

Larger flow magnetization $\sigma$ results in lower accretion rate $\dot{M}$. Larger magnetic field and turbulent velocity field
exerts larger back-reaction force on matter. Also, transformation of gravitational energy into thermal happens more readily if
magnetization is larger. Larger thermal energy means larger gas pressure and larger back-reaction force on matter striving to
fall onto the central object.

Several factors lead to higher magnetization. Larger outer magnetization $\sigma_\infty$ makes magnetization in the entire flow
$\sigma$ larger. Then larger dissipation length scale $\gamma$ allows for smaller dissipation of magnetic field. Larger magnetic
helicity $\xi$ also lowers magnetic energy dissipation and leads to larger magnetization $\sigma$. These correlations can be
observed on Figure~\ref{fig2}. Increase of the relative length scale of energy containing eddies $\gamma$ from $0.2$ to $2$
results (Fig.~\ref{fig2}a) in about 2 times drop in accretion rate $\dot{M}.$ Accretion rate stays constant (Fig.~\ref{fig2}b) at
small values of outer magnetic helicity $\xi_\infty.$ However, $\dot{M}$ drops an order of magnitude as turbulence approaches
highly helical state at outer boundary $3r_B$ with $\xi_\infty$ close to $0.5.$ The dependence of $\dot{M}$ on outer
magnetization $\sigma_\infty$ is not quite steep: accretion rate gradually decreases about $4$ times as outer magnetization
increases $3$ orders of magnitude from $0.001$ to $1.$ Surprisingly, accretion rate does not rise to $\dot{M}_B$
(Fig.~\ref{fig2}c) even for very small outer magnetization $\sigma_\infty\sim0.001$ for non-relativistic equation of state. Even
small outer magnetic field increases inwards and influences flow dynamics.

Accretion rate is systematically about $40\%$ higher (Fig.~\ref{fig2}) for relativistic equation of state (solid line) compared
to non-relativistic equation of state (dashed line), because magnetized system has some properties of a non-magnetized one.
Formula for Bondi mass accretion rate (\ref{bondi_rate}) is valid only for non-relativistic monatomic gas that has an adiabatic
index $\Gamma=5/3.$ Accretion rate is higher for lower $\Gamma$ and is about $3$ times larger \citep{shapiro} in case of
ultrarelativistic particles with adiabatic index $\Gamma=4/3.$ Accretion rate $\dot{M}$ is determined by relation (\ref{maxrate})
at a sonic radius $r_s$ that is smaller than $10^3r_{\rm g}$ (Fig.~\ref{fig2}d). Electrons become relativistic at somewhat larger
radius about $10^3r_{\rm g}$ in the solutions of system (\ref{system}). This leads to gas adiabatic index $\Gamma$ (magnetic field is disregarded)
lower than $5/3$ at sonic point $r=r_s.$ Thus accretion rate is considerably larger in case of relativistic equation of state.

It is also instructive to trace the dependence of sonic radius $r_s$ on parameters.  Sonic radius for hydrodynamic accretion of
non-relativistic monatomic gas is equal to several Schwarzschild radii $r_s=2\div10 r_{\rm g}$ \citep{pidoprygora}. Sonic radius
is a considerable fraction of $r_B$ for a gas with adiabatic index $\Gamma$ substantially smaller than $5/3$ for non-magnetized
accretion \citep{bondi52}. Magnetized accretion has the same properties. Non-relativistic EOS (solid line) results in very small
sonic radius $r_s=7\div11r_{\rm g}$ (Fig.~\ref{fig2}d). Sonic radius for relativistic EOS (dashed line) is $r_s=300\div1200r_{\rm
g}$ about the radius where electrons become relativistic $r\sim 10^3r_{\rm g}.$ The value of sonic radius drops several times as
plasma outer magnetization $\sigma_\infty$ increases from $0.001$ to $1.$ As outer magnetization $\sigma_\infty$ increases,
accretion rate drops (Fig.~\ref{fig2}c), because density $\rho$ and gas inflow speed $v$ decrease. Then effective sound speed
$V_s$ equals the inflow speed $v$ at a point closer to the black hole.

\begin{figure}[h]
\epsscale{.85} \plotone{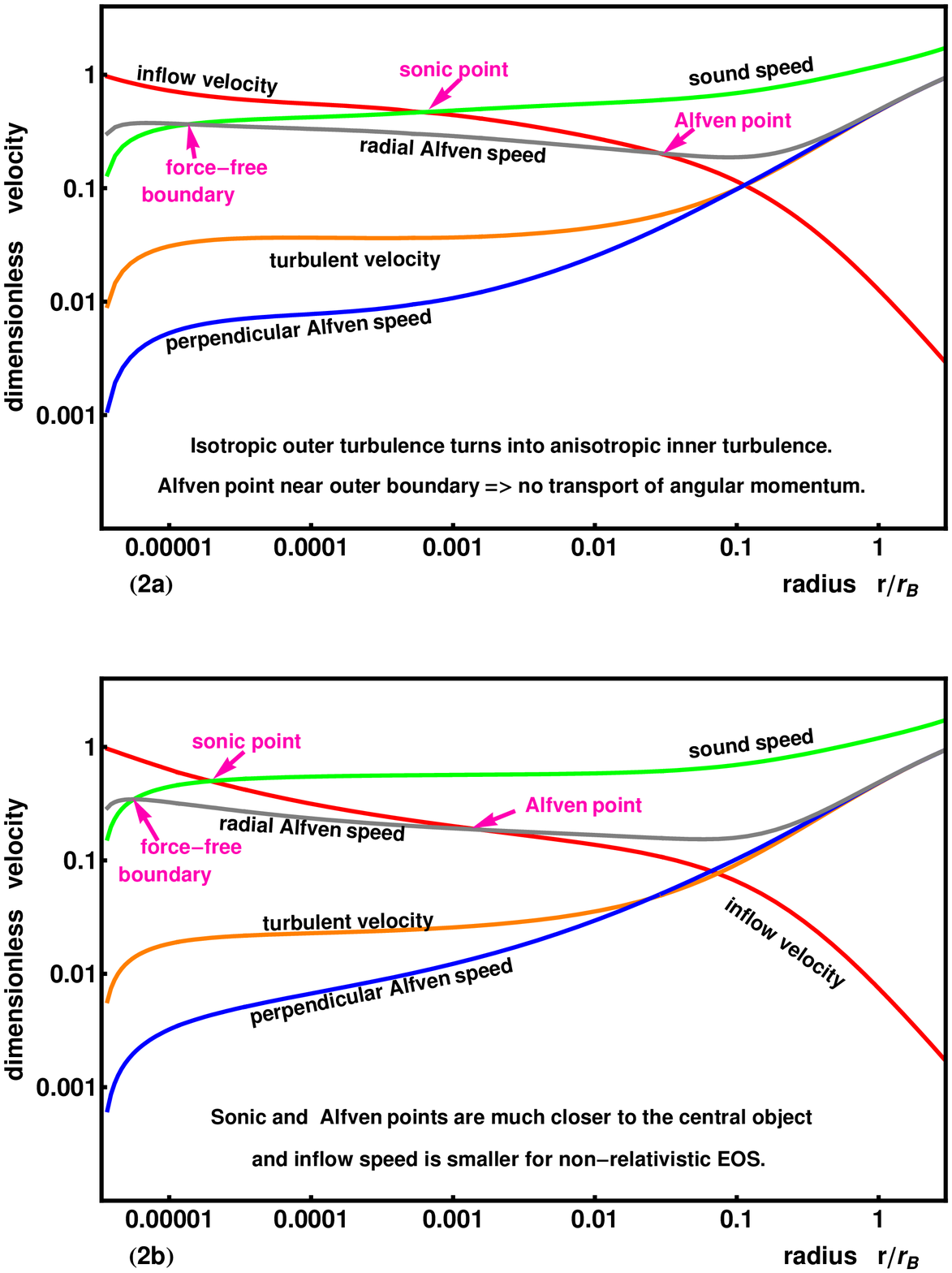}
 \caption{Flow velocities, normalized to free-fall speed  verses radius for \textbf{maximum-rate
 solution}: sound speed, inflow velocity, radial Alfven speed, 1-D perpendicular Alfven speed, turbulent velocity. Parameters $\sigma_\infty=1$,
$\gamma=1$, $\xi_\infty=0.025.$ Relativistic 1-T equation of state is on Figure~\ref{fig3}a, non-relativistic 1-T EOS is on
Figure~\ref{fig3}b.}
 \label{fig3}
\end{figure}

Inflow velocity $v$ as well as other characteristic velocities of the flow are depicted on Figure~\ref{fig3} as functions of
radius $r$ for the reference model with $\sigma_\infty=1,$ $\gamma=1,$ $\xi_\infty=0.025$. All velocities are normalized to the
free-fall speed
\begin{equation}\label{v_ff}
v_{ff}=c\sqrt{\frac{r_{\rm g}}{r-r_{\rm g}}}.
\end{equation} I also normalize perpendicular Alfven velocity $v_{A\perp}$
and turbulent speed $u$ to one dimension. Horizontal line on
Figure~\ref{fig3} corresponds to radial dependence $r^{-1/2}.$

Inflow velocity $v$ monotonically increases inwards, whereas sound speed $c_s$ monotonically decreases with intersection almost
at the sonic point. Radial Alfven velocity $v_{Ar},$ perpendicular Alfven velocity $v_{A\perp}$ and turbulent velocity $u$
(Fig.~\ref{fig3}) start out as constants from the outer boundary at $3r_B,$ where turbulence is sustained by external pumping.
Then these velocities increase and deviate from one another. Radial Alfven velocity $v_{Ar}$ appears to be much larger than
$v_{A\perp}$ and $u$ in the inner accretion region. This fulfills the expectations of earlier models
\citep{shakura,scharlemann,beskin}. At small radius turbulence is driven by freezing-in amplification of magnetic field and
random velocity. Left-hand sides of turbulence evolution equations (\ref{B_r1}), (\ref{B_perp_r1}), and (\ref{u_r1}) dominate
over corresponding terms with external driving for radius $r\lesssim 10^4r_{\rm g}.$ Internal driving of $v_{Ar}$ is much more
effective than driving of $v_{A\perp}$ and $u$. Therefore radial Alfven velocity $v_{Ar}$ is larger than other two speeds. This
refutes any model with isotropic magnetic field.

\begin{figure}[h]
\plotone{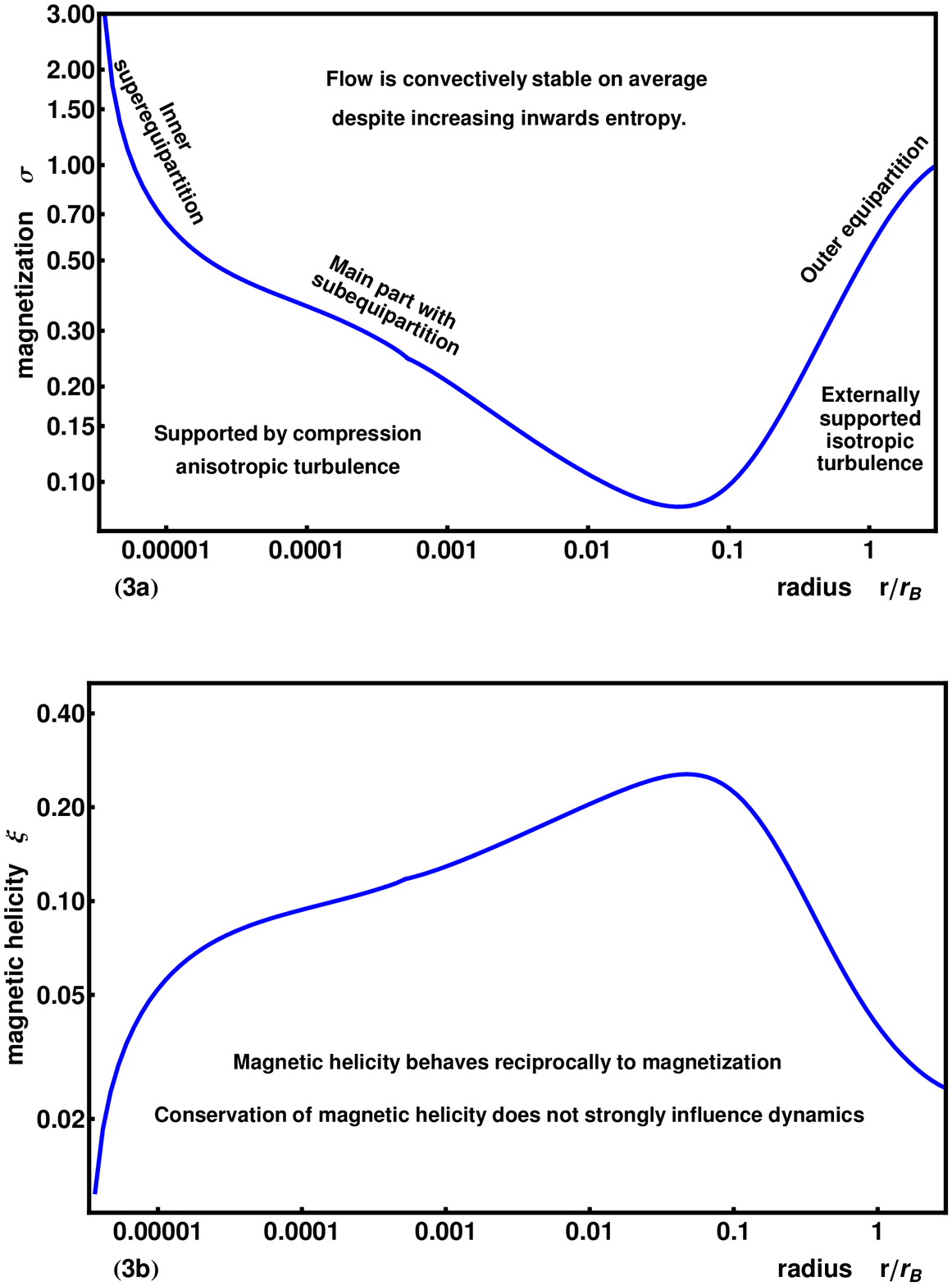} \epsscale{1}
 \caption{Magnetization $\sigma$ verses dimensionless distance from the compact object $r/r_B$ is on Figure~\ref{figmagn}a.
Dimensionless magnetic helicity $\xi$ verses dimensionless distance from the compact object $r/r_B$ is  on Figure~\ref{figmagn}b.
Both are for the \textbf{maximum-rate solution} with relativistic equation of state.}
 \label{figmagn}
\end{figure}

Several pairs of lines intersect on velocity plot (Fig.~\ref{fig3}). I consider three main intersection points for the reference
model with $\sigma_\infty=1,$ $\gamma=1,$ $\xi_\infty=0.025,$ and relativistic EOS (Fig.~\ref{fig3}a). Crossing of inflow
velocity $v$ and sound speed $c_s$ occurs almost at the sonic point at $r_s,$ determined by relation (\ref{maxrate}) with
critical velocity $V_s$ (eq. [\ref{sonic}]). No plasma waves can escape from within the region with high inflow velocity $v>V_s.$
Approximately $c_s\approx V_s$ at sonic point $r_s\approx6\times10^{-4}r_B$, because of low magnetization $\sigma\approx 20\%$ in
that region (Fig.~\ref{figmagn}a). Alfven point is determined by equality $v=v_{Ar}$ at radius $r_A.$ Alfven waves cannot escape
from within the region where inflow speed is greater than radial Alfven speed $v_{Ar}.$ Equality holds at relatively large radius
$r_A\approx0.03r_B.$ The third combination of the same three velocities also gives a characteristic intersection point. Radial
Alfven speed $v_{Ar}$ increases faster inwards and becomes equal to sound speed $c_s$ at about $r\approx4r_{\rm g}.$ Further
relative increase of $v_{Ar}$ leads to magnetic energy dominated flow, what can be traced on magnetization plot
(Fig.~\ref{figmagn}a).

Figure~\ref{figmagn}a shows evolution of plasma magnetization $\sigma$ with radius $r$ for the reference model. Thermal energy
equipartition assumption does not hold, id est turbulent energy does not equal to constant fraction of thermal energy
$\sigma\ne\rm const.$ Magnetization $\sigma$ varies more than one order in magnitude from $0.07$ to $3.$ It starts out at initial
$\sigma_\infty=1$ at $3r_B,$ where turbulence is supported by external energy input $Q_+=\rm const.$ Then $\sigma$ deviates down
as $r$ decreases. Magnetization $\sigma$ drops, because length scale $L$ decreases with radius $r$ that causes turbulence to
decay faster. At about $0.03r_B$ magnetization starts to rise as internal turbulence driving takes over. Inflow velocity $v$
slightly deviates up from Alfven velocity $v_A$ as $r$ decreases. Since internal driving rate is proportional to $v$ (left-hand
sides of equations (\ref{B_r1}), (\ref{B_perp_r1}), and (\ref{u_r1}) dissipation rate is proportional $v_{Ar}$, parameter
$\sigma$ grows slightly with decreasing radius. The growth is about a factor of $5$ for $3.5$ orders of magnitude decrease in
radius. Magnetization $\sigma$ jumps up in the region very close to the event horizon of the black hole. However, this jump may
originate from inconsistent treatment of General Relativity.

The dependence of magnetic helicity $\xi$ on radius is shown on Figure~\ref{figmagn}b. Helicity $\xi$ behaves almost reciprocally
to magnetization $\sigma$ from Figure~\ref{figmagn}a. Such a behavior can be seen from magnetic helicity equation (\ref{hel_r1}).
Magnetization $\sigma$ decreases order of magnitude during the transition from externally supported to internally supported
turbulence around $r\approx0.03r_B.$  Magnetic helicity $\xi$ also increases an order of magnitude from $0.025$ to $0.2.$ Then
$\xi$ gradually decreases down to initial value. Thus magnetic helicity $\xi$ does not change dynamics if it is initially small
$\xi_\infty\lesssim0.1$. Only when $\xi_\infty$ is large, accretion rate drops.

Deviation of inflow velocity $v$ from the free-fall scaling $r^{-1/2}$ makes a density profile in magnetized flow different from
that in standard Advection Dominated Accretion Flow (ADAF). I consider the flow where energy is only advected inward.
Nevertheless, I obtain
\begin{equation}\label{rho_profile} \rho\propto
r^{-\zeta} \quad{\rm with}\quad \zeta\approx 1.25\end{equation}
almost independently on the parameters or the equation of state,
somewhat shallower than $\rho\propto r^{-1.5}$ in ADAF.

The only question left is how well this flow with maximum accretion rate can describe the real situation with large angular
momentum $l$. Given the solution of the system (\ref{system}) I can check whether the condition for effective angular momentum
transport condition (\ref{condition}) holds. Condition (\ref{condition}) breaks when evaluated for maximum-rate solution with
parameters $\xi_\infty,$ $\sigma_\infty,$ and $\gamma$ within the chosen ranges and circularization radius $r_{\rm cir}>r_{\rm
g}.$ This means a flow with maximum accretion rate is unable to effectively transport the angular momentum outward. The same
conclusion can be made simpler. The transport of angular momentum is a magnetic process. So, $l$ can be transported only by
Alfven waves. However, Alfven waves cannot escape from the region within $r_A\approx 0.03r_B$ from the compact object that makes
angular momentum transport impossible even from quite large radius.

\subsection{Solution with Effective Angular Momentum Transport}\label{physical_solution}
Solution with large outer angular momentum $l\gg r_{\rm g} c$ and small viscosity may have properties, substantially
different from those of maximum-rate solution. The actual details of the solution and allowed accretion rate depend on
how this angular momentum is transported. For the simple estimate I suppose that the accretion rate is determined by
the equality in angular momentum transport condition (\ref{condition}). Maximum accretion rate $\dot{M}$ for condition
(\ref{condition}) appears to be about two orders of magnitude lower than Bondi rate $\dot{M}_B$ (eq.
[\ref{bondi_rate}]).

\begin{figure}[h]
\plotone{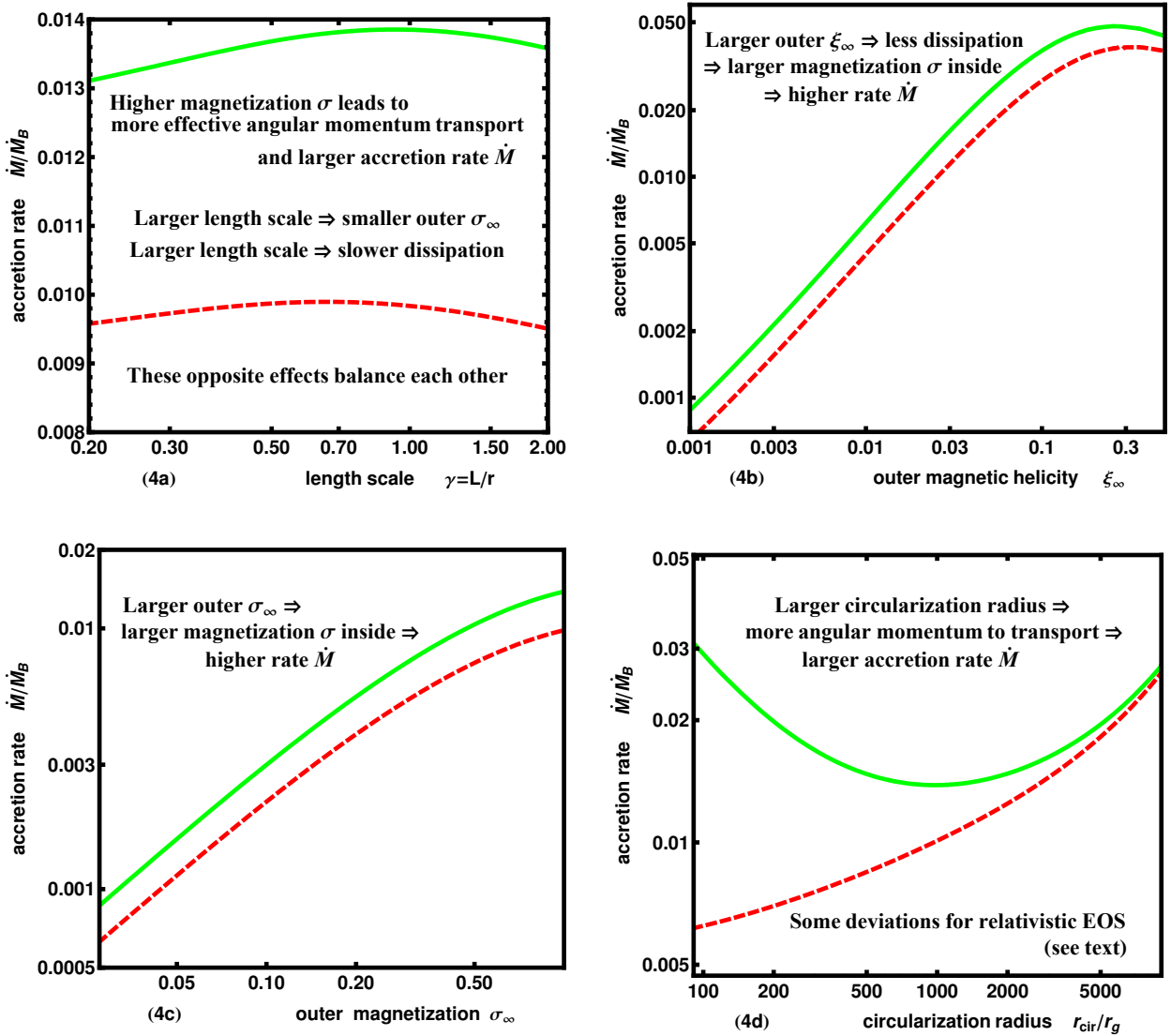}
 \caption{\textbf{Solution with angular momentum transport.} Dependence of the accretion rate in units of Bondi rate
 on dimensionless parameters:
characteristic length scale $\gamma$ (Fig.~\ref{fig4}a), outer magnetic helicity $\xi_\infty$ (Fig.~\ref{fig4}b), outer
magnetization $\sigma_\infty$ (Fig.~\ref{fig4}c), and circularization radius $r_{\rm cir}$ in units of $r_{\rm g}$
(Fig.~\ref{fig4}d). I take the reference model to have the following parameters: $\gamma=1$, $\sigma_\infty=1$, $r_{\rm
cir}=10^3r_{\rm g}$, $\xi_\infty=0.025$. Non-relativistic 1-T equation of state (dashed) verses relativistic 1-T equation of
state (solid).}
 \label{fig4}
\end{figure}

 I add one parameter in modelling: unknown 
 circularization radius $r_{\rm cir}$ for specific angular momentum $l$ (eq. [\ref{l}]). I take it to be $r_{\rm cir}=10^3r_{\rm
g}$ for the reference model. Plots of the accretion rate verses model parameters are shown on Figure~\ref{fig4}. Dependencies for
the rotating solution (Fig.~\ref{fig4}) have the opposite slopes to those for the maximum-rate solution on Figure~\ref{fig2}.
Accretion rate $\dot{M}$ increases with increasing outer magnetization $\sigma_\infty$ (Fig.~\ref{fig4}b) and increasing outer
magnetic helicity $\xi_\infty$ (Fig.~\ref{fig4}c). Both effects lead to higher plasma magnetization $\sigma$. I showed in the
previous subsection~\ref{maximum_rate_solution} that the magnetic field plays an inhibiting role on matter inflow, and that the
larger the magnetic field is, the smaller the accretion rate $\dot{M}$ is. However, the correlation between the magnetic field
and accretion rate is the opposite in case of the rotating flow. Accretion rate quantitatively agrees with relation 
for ADAF flows $\dot{M}\sim \alpha \dot{M}_B\sim\sigma\chi \dot{M}_B$ \citep{narayan_adaf} with $\sigma\sim0.01$ at $r_{\rm cir}$ (Fig.~\ref{figang}a).

The allowed by condition (\ref{condition}) inflow speed $v$ is proportional to the product of radial Alfven speed $v_{Ar}$ and
perpendicular Alfven speed $v_{A\perp}.$ Larger magnetic field results in larger transport of angular momentum outward, so larger
inflow velocity $v$ and larger accretion rate are possible. Larger outer magnetization $\sigma_\infty$ and larger outer magnetic
helicity $\xi_\infty$ both lead to higher magnetization $\sigma$ and higher magnetic field. Inhibiting effect of magnetic field
is smaller in case of lower accretion rates $\dot{M}$ and lower inflow velocities $v.$ Lower $v$ results in lower relative
driving of turbulence that makes magnetic field weaker. Weaker magnetic field has weaker influence on dynamics. In sum, larger
magnetic field $B$ results in larger accretion rate $\dot{M},$ when it needs to transfer angular momentum.

The dependence of $\dot{M}$ on length scale $\gamma$ is obscured by the dependence of external driving on $\gamma$. Accretion
rate $\dot{M}$ is smaller for smaller magnetic field, but the state of low magnetization can be achieved in two different ways.
Firstly, magnetic field decays faster when $L$ decreases. However, the plasma at circularization radius $r_{\rm cir}=10^3r_{\rm
g}$ is still partially influenced by the outer boundary conditions. Internal driving does not depend on $L,$ whereas external
driving is stronger and magnetization $\sigma$ is higher, when $L$ is small. The described two effects balance each other and
make accretion rate $\dot{M}$ almost independent of dimensionless length scale $\gamma$ (Fig.~\ref{fig4}a).

Accretion rate $\dot{M}$ decreases with the decrease of circularization radius $r_{\rm cir}$ (Fig.~\ref{fig4}d) for
non-relativistic equation of state. To explain this, I trace on Figure~\ref{fig5}b all the quantities that enter angular momentum
transport condition (\ref{condition_b}) for the reference model. Velocities normalized by the free-fall speed (eq. [\ref{v_ff}])
are shown on Figure~\ref{fig5}b. Inflow speed $v$ and radial Alfven velocity $v_{Ar}$ reach free-fall scaling at about $0.02r_B.$
Only perpendicular Alfven velocity $v_{A\perp}$ has a different dependence on distance from the central object for $r<0.02r_B.$
Because $v_{A\perp}$ decreases with radius, the allowed $v$ and $\dot{M}$ are smaller for smaller circularization radius.

However, the accretion rate increases for small circularization radii for 1-T equation of state (Fig.~\ref{fig4}d, solid line).
This is the consequence of the decreasing gas adiabatic index, when electrons reach relativistic temperatures. Solutions with
lower adiabatic index are known to have larger accretion rates \citep{bondi52} that is equivalent to the lower inflow speeds $v$
in the solutions for the fixed matter inflow rate. Velocity $v$ (Fig.~\ref{fig5}a) starts deviating down from the self-similar
$r^{-1/2}$ solution at approximately $10^3r_{\rm g},$ making the solutions with higher $\dot{M}$ possible. In fact, condition
(\ref{condition}) for the solutions with small $r_{\rm cir}$ becomes critical at some fixed point $r_d>r_{\rm cir}$ instead of
reaching equality at $r_{\rm cir}$ (eq. [\ref{condition}b]). Therefore, according to condition (\ref{condition}a), maximum value
of the inflow speed grows with the decrease of circularization radius as $v\propto r_{\rm cir}^{-1/2},$ explaining the rise of
accretion rate for small $r_{\rm cir}$ (Fig.~\ref{fig4}d, solid line) for 1-T equation of state.

Solution for non-relativistic equation of state, in turn, possess its own feature. Self-similar flow (see
Appendix~\ref{self_similar_solution}) settles in at $10^3r_{\rm g},$ making accretion rate almost independent on circularization
radius (Fig.~\ref{fig4}d). Magnetic helicity $\xi$ in such a flow is a number about unity what is consistent with self-similar
solution obtained in Appendix~\ref{self_similar_solution}. Self-similar flow can not establish for 1-T equation of state, because
relativistic effects become important before it establishes and break self-similarity.

In fact, magnetization $\sigma$ and magnetic helicity $\xi$ (Fig.~\ref{figang}) are not constant at small radii for correct 1-T
EOS, because these relativistic corrections work. At about $0.01r_B$ magnetization reaches almost constant level
$\sigma\approx0.02$ (Fig.~\ref{figang}a) and then starts to slightly deviate down, because equilibrium $\sigma$ for matter with
lower gas adiabatic index $\Gamma<5/3$ is lower. Magnetic helicity $\xi$ behaves (Fig.~\ref{figang}b) the opposite way to
magnetization $\sigma:$ magnetic helicity reaches $\xi\approx1.5$ at $0.01r_B$ and starts to slightly deviate up as the radius
decreases.

\section{DISCUSSION OF THE MODEL}\label{sec_discussion}
I present the sophisticated analytical model to determine the properties of spherical magnetized accretion. The common
assumptions of magnetic field isotropy and thermal equipartition are released, but many assumptions are still left. As usually in
fluid dynamics a lot of simplifications are made during the course of elaboration. The validity of almost everything can be
questioned. The system of equations (\ref{system}) may not describe the real flow (subsection~\ref{subsection_flow}) or may have
some inaccuracies (subsection~\ref{subsection_magnetic_field}). Gas cooling may not be neglected
(subsections~\ref{subsection_cooling}). Convection and diffusion may change the flow structure (subsection~\ref{subsection_convection}).
The equation of state was also found to influence the dynamics (subsection~\ref{subsection_EOS}). Let me discuss all these topics and
determine the practical significance of the model.

\subsection{Real Flow}\label{subsection_flow}
Presented model is partially applicable to the real systems. It may describe some gas flows onto Supermassive Black Holes in Low
Luminosity Galactic Centers, in particular in the center of our Galaxy. These flows are geometrically thick \citep{narayan} and
may have low angular momentum \citep{monika}. However, the real flows may have properties that my model cannot handle in its
current state. First of all, the sources of matter and external driving should be explicitly accounted for. Secondly, the
self-consistent angular momentum transport theory is needed.

The material is mainly supplied to the central parsec of the Milky Way by stellar winds \citep{quataert_wind}. The wind-producing
stars have a broken power-law distribution as a function of radius \citep{baganoff}. Some stars are as close to the central black
hole as $0.1r_B$ \citep{Ghez}. The stars supply too much material to be accreted, therefore there exist an outflow
\citep{quataert_wind}. Bondi radius coincides with the radius where inflow starts to dominate outflow in numerical simulations
with the accretion rate $\dot{M}\sim10^{-6}M_\odot {\rm year}^{-1}$ \citep{cuadra}. Maximum accretion rate in the solution with
zero angular momentum is $0.2\dot{M}_B\approx10^{-6}M_\odot{\rm year}^{-1}$ and $0.01\dot{M}_B$ for the rotating flow. So that
the transition from the outflow to the inflow happens at $r\gtrsim10^5r_{\rm g}.$

I can show that outflow from $r\gtrsim10^5r_{\rm g}$ does not change the accretion rate from calculated. Outflows substantially
alter the value and the sign of inflow velocity $v$ in the system (\ref{system}). However, the differences in inflow velocity do
not influence any other quantity as long as three conditions are satisfied:
\begin{enumerate}\label{inflow_outflow}
\item{$v$ is much smaller than gas particles velocity $v_p,$ bulk kinetic energy of gas is negligible in the outflow region,}
\item{external driving of turbulence $Q_+$ dominates over internal driving there,} \item{condition on $\dot{M}$ is set in the
inflow region.}
\end{enumerate}
The first two conditions are satisfied down to $r\sim10^4r_{\rm g}$ (Fig.~\ref{fig3} and Fig.~\ref{fig5}). The third condition
holds for maximum rate solution, because condition on $\dot{M}$ is set at the sonic point about $10^3r_{\rm g}$ from the central
object. It also hold for the solution with angular momentum transport, because the condition on $\dot{M}$ is usually set at the
inner boundary $10^3\div10^4r_{\rm g}.$ All three above conditions hold, hence outflows of stellar winds do not substantially
change the accretion rate or any quantity in the system.

\subsection{Treatment of Magnetic Field}\label{subsection_magnetic_field}
The long history of accretion theory has many accepted models based on ideas, extended beyond the area of applicability of these
ideas. For example, general relativity was substituted with Paczynski-Wiita gravitational potential \citep{paczhynski,shakura}.
Magnetic field was long treated similar to the normal matter \citep{narayan,coker}. Displacement current was neglected in
magnetic field dynamics that allowed to treat magnetic field without electric field \citep{scharlemann}. System of viscous
equations describe viscosity by a single parameter \citep{shakura,landau6,landau8,biskamp03}. Gyrokinetics is used to solve the
problems with non-Maxwellian distribution functions \citep{sharma}, power-law non-thermal electrons are usually present in plasma
\citep{yuan_jet}.

Described above model is extended in several ways, mainly with regard to magnetic field. Isotropic MHD system of turbulent
equations (\ref{iso_turbulence}) describes the real box collisional turbulence quite well, because it corresponds to convergent
set of simulations. Collisionality assumes that medium behaves like many particles with short-range interactions. However,
astrophysical medium of interest is always collisionless with prevailing long-range interactions. I inconsistently use the
results of numerical simulations of collisional MHD (eqs. [\ref{gen_continuity}-\ref{gen_solenoid}]) with magnetic resistivity
$\nu_M$ on the order of viscosity $\nu,$ because the realistic simulations of collisionless plasma turbulence are not done and
are unlikely to be done in the near future \citep{scheko}.

Observations of astrophysical turbulence may give more information than numerical simulations. A special case of collisionless
plasma is plasma with random kinetic energy much smaller than random magnetic energy. This regime is a good picture of Sun corona
with all plasma effects into play \citep{Aschwanden}. Dissipation of magnetic loops with low kinetic energy proceeds mainly via
reconnections. The timescale of reconnective dissipation was found to be \begin{equation}\label{tau_rec}\tau_{\rm
rec}\approx20\frac{L}{v_A}\end{equation} in solar flares \citep{noglik}. The same number was also predicted by
\citet{vishniac_rec}. Collisional MHD turbulence has much smaller dissipation timescale
\begin{equation}\label{tau_diss}\tau_{\rm
diss}\approx1\frac{L}{v_A}\end{equation} (eqs. [\ref{iso_turbulence}, \ref{coeff_hats}]). Plasma has large kinetic energy in the
outer region of accretion flow, where turbulence is externally supported. Timescale $\tau_{\rm diss}$ (eq. [\ref{tau_diss}]) may
be appropriate there. Kinetic energy $E_K$ decreases to smaller radii and magnetization $\sigma$ increases (Fig.~\ref{fig3}) in
case of zero angular momentum (\ref{maximum_rate_solution}). Accretion flow there may resembles solar Corona \citep{Aschwanden}.
Dissipation timescale may increase order of magnitude and be close to $\tau_{\rm rec}$ (eq. [\ref{tau_rec}]). This increase would
lead to much lower accretion rate, because higher magnetic field leads to lower $\dot{M}.$ Matter infall may eventually proceed
through channels of lower magnetic field \citep{igumen06}.

Even if I assume that box isotropic turbulent system of equations (\ref{iso_turbulence}) with coefficients (\ref{coeff_hats}) is
applicable to isotropic turbulence, there are at least four complications in building the full anisotropic theory.

First of all, I need to introduce arbitrary coefficients $c_{uB2},$ $c_{BB2},$ $c_{Bu2}$ to describe isotropization of
anisotropic magnetic field and anisotropic energy transfer between magnetic field and fluid motions. Reasonable values of these
coefficients were taken to satisfy rather loose analytical tests (Appendix~\ref{section_analytic_tests}). However, changes in
these coefficients do not lead to dramatically different accretion rate or flow structure. Setting $c_{BB2}=c_{BB1}$ instead of
$c_{BB2}=0$ leads to only $10\%$ of $\dot{M}$ change for the reference model. All seven introduced coefficients $c_{xx}$ may
themselves depend on anisotropy of the magnetic field. The details of anisotropic MHD are still debatable
\citep{goldreich,boldyrev_turbo}. I leave the incorporation of anisotropic MHD model into accretion theory for future work.

Secondly, the presented theory is not general relativistic. Accretion rate $\dot{M}$ appears to be insensitive to the choice of
gravitational potential. The condition on $\dot{M}$ is set at about $10^3r_{\rm g}$ in case of relativistic EOS and zero angular
momentum $l$. Sonic point is situated close to the black hole at $r_s=5\div10r_{\rm g}$ for non-relativistic equation of state.
But $1\%$ increase of $\dot{M}$ leads to the sonic point at $r_s>100r_{\rm g},$ independent of the way to mimic general
relativity. However, the region near the black hole is important, because part of synchrotron IR radiation as well as part of
radio emission comes from several Schwarzschild radii \citep{narayan98,falcke,marrone}. Thus, to fully constrain theory by
observations general relativistic magnetohydrodynamics is a must.

In third, magnetic helicity $H$ involves numerous complications. Magnetic helicity evolves in the region that is frozen into
matter. The distance $L_{||}$ between radial boundaries of this region is proportional to inflow velocity $v,$ thus $L_{||}$
increases with increasing $v$ and at some point $L_{||}>r,$ whereas size in the angular direction is about $L=\gamma r.$ A part
of the region is getting sucked into the black hole, while a part is still situated at fairly large radius $r$. Equation of
magnetic helicity evolution (\ref{hel_r1}) holds only if I assume even redistribution of magnetic helicity over the mass of
plasma. This holds for frozen magnetic field, but in reality diffusion and convection are present. Diffusion may change the
results for $H$ (eq. [\ref{hel_r1}]) as well as for the entire flow pattern. I also leave these uncertainties for future
research.

In fourth, it was recently suggested by \citet{beskin} that ions and electrons should be viewed in accretion as confined by
magnetic field lines. This is the opposite of standard picture where magnetic field lines are frozen into matter
\citep{scharlemann}. The former case has higher heating rate of matter under contraction \citep{beskin}, because of conservation
of the first adiabatic invariant  $I=3c p_t^2/(2 e B)=\rm const$ \citep{landau2}. Here $p_t$ is a particles momentum in the
direction perpendicular to ${\bf B}.$ However, only highly magnetized flows with magnetization $\sigma>1$ conserve $I.$
Non-linear collective interactions of particles in low-$\sigma$ plasma are likely to isotropize their distribution. When
particles are heated isotropically under contraction, general Magneto-Hydrodynamics (eqs.
[\ref{gen_continuity}-\ref{gen_solenoid}]) works \citep{landau8} and heating rate stays unchanged. Magnetization in computed
models is below unity (Fig.~\ref{figmagn}a and Fig.~\ref{figang}a). Thus application of first adiabatic invariant conservation to
magnetized accretion flow seems irrelevant.

Finally, mean rotation of the flow also creates anisotropy. Because the inner gas rotates faster than the outer,
MagnetoRotational Instability (MRI) works. It produces the additional driving of magnetic field that may be concurrent to other
sources. MRI \citep{hawley} has a timescale
\begin{equation}\label{tau_MRI}
\tau_{MRI}=-\left(r \frac{d (l/r^2)}{dr}\right)^{-1}.
\end{equation}
When MRI timescale becomes larger then dynamic timescale $\tau_{\rm dyn}=r/v,$ field amplification occurs mainly because of
regular shear tangential motion, instead of regular radial motion. MRI may be crucial even in the region without rotational
support. Full consideration of effects of angular momentum on the flow is the subject of the next study.

\subsection{Radiative Cooling}\label{subsection_cooling}
The system of equations (\ref{system}) describes the accretion flow, where all the energy is stored in the same piece of matter
where it initially was. There is no energy loss by diffusive or radiative cooling. But whether such a model is realistic.

Let me estimate the radiative cooling first. Line cooling is more effective than bremsstrahlung cooling for temperatures about
$T_\infty\approx1.5\times10^7K.$ Line cooling function is $\Lambda\approx6\times10^{-23}n^2(T/10^7K)^{-0.7}$ ${\rm erg}~{\rm
cm}^{-3}~{\rm s}^{-1}$ \citep{sutherland}. Thus characteristic cooling time $\tau_{\rm cool}$ is
\begin{equation}\label{time_cool}
\tau_{\rm cool}=\frac{3R T \rho}{2\Lambda \mu}\approx
1\times10^{12}{\rm s}
\end{equation} for our Galactic Center accretion.
The dynamic timescale $\tau_{\rm dyn}=r/v$ for accretion with rate
$\dot{M}=0.1\dot{M}_B$ (eq. [\ref{bondi_rate}]) is
\begin{equation}\label{time_dyn}
\tau_{\rm dyn}=\frac{\rho r^3}{\dot{M}}\approx5\times10^{10}{\rm
s}
\end{equation}
with continuity equation (\ref{mdot_r}) at radius $r=r_B$ (eq. [\ref{bondi}]). Cooling time is about $20$ times larger than
inflow time in the region where outflows dominate. Nevertheless, anisotropy of stellar winds may lead to significant cooling of
some clumps of matter \citep{cuadra_cool}. Even the disk may form \citep{cuadra}. Careful calculation with line cooling is yet to
be done.

\subsection{Convection \& Diffusion}\label{subsection_convection}
The system (\ref{system}) does not include diffusive or convective transport of quantities. Thus the system represents Advection-Dominated flow, where
magnetic field and gas can exchange energy between each other. The exact model would include transport of momentum, energy, magnetic field, 
magnetic helicity that may or may not influence the dynamics.

First or all, any type of convective or diffusive motion would happen at a speed $v_c$ not exceeding the maximum of turbulent speeds, radial Alfven speed $v_c<v_{Ar}.$ This leads to the transition from convection dominated to advection dominated flow at several dozens $r_{\rm g}$ in the case with rotation \citep{abr}. Correspondingly, inflow speed $v$ becomes large $v_c\sim v$ \citep{popham}. Transport becomes ineffective at $r\lesssim r_A,$ where  $r_A$ is the radius of Alfven point. According to Fig.~\ref{fig3}a, Alfven point in my spherical solutions lies at $r_A\sim0.03r_B.$ Thus diffusion and convection are strongly suppressed in the inner flow. By the same reason, magneto-thermal instability (MTI)\citep{parrish} is not supposed to play any role for spherical inflow, but may play a role in a case with rotation. For the non-conductive convective stability criterion see Appendix~\ref{section_convection}.

However, speed of electrons $v_e$ may overcome the speed of sound $c_s,$ so electron conduction may in principle transport energy from within $r_A$ \citep{johnson}. It is yet unclear whether electron conduction is suppressed at high inflow velocity $v>v_{Ar},$ because electrons may be bound to the field lines of tangled magnetic field. The efficiency of conduction is a free parameter. If efficiency is close to maximum and conduction is not inhibited, then accretion rate may be $1\div2$ orders of magnitude lower than Bondi rate $\dot{M}_B$ \citep{johnson}, thus accretion rate would be limited by conduction and not by backreaction of the magnetic field. Other types of energy transport \citep{parrish} may kick in for lower accretion rates. The correct calculation with magnetic field and better prescription for conductivity is yet to be done.

\subsection{Equation of State}\label{subsection_EOS}
The difference in accretion rate $\dot{M}$ between one-temperature relativistic and 1-T non-relativistic EOSs is up to $40\%$ for
maximum-rate solution (subsection~\ref{maximum_rate_solution}) and up to several times for solution with effective angular
momentum transport (subsection~\ref{physical_solution}). Solution with smaller gas adiabatic index $\Gamma$ has larger accretion
rate $\dot{M}$ \citep{shapiro}. Gas adiabatic index gradually falls from $\Gamma=5/3$ to $\Gamma=1.43$ in case of relativistic
EOS as matter approaches the black hole.

However, the electron temperature $T_e$ is unlikely to be equal to
ion temperature $T_i.$ Electron temperature $T_e$ is usually
modelled to be lower than $T_i$ \citep{narayan}. This
two-temperature model has lower gas pressure support and larger
gas adiabatic index $\Gamma$ than 1-T model with $T=T_i.$ Lower
gas pressure leads to higher accretion rate, larger $\Gamma$ leads
to lower accretion rate. The combination of these two effects is
expected to change the accretion rate by about the same $40\%$ as
between relativistic and non-relativistic 1-T EOSs. The exact
details depend on the two-temperature model chosen.

\section{OBSERVATIONS}\label{sec_observations}
Proposed quasi-spherical magnetized accretion model is aimed to explain plasma flow onto SuperMassive Black Hole Sgr A* in our
Galactic Center. Many observations of this source are made. These observations reasonably agree with the results of my model.

A common misconception about Chandra X-Ray observations of Sgr A* exists in literature. X-Rays mainly originate in the region
that lies further than Bondi radius $r_B$ from the central object. Thus characteristic density $\rho_\infty$ and temperature
$T_\infty$ far from the Black Hole can be found \citep{baganoff}. If one knows the mass $M$, this automatically gives Bondi
accretion rate $\dot{M}_B$ (eq. [\ref{bondi_rate}]). However, accretion rate is not necessarily determined by this formula
(\ref{bondi_rate}), unlike some papers suggest \citep{bower}. In my model accretion rate $\dot{M}$ is independent on radius and
is smaller than $\dot{M}_B.$

IR \citep{eckart} and Radio \citep{shen} observations are difficult to interpret, because fluxes in these diapasons
depend strongly on the accretion model. Density of matter $\rho$ is better constrained by observations than accretion
rate $\dot{M}.$ The general agreement \citep{yuan_jet} is that density $\rho$ should be lower than in Bondi solution
$\rho_B$ in the region close to the black hole. Solutions with outflows \citep{yuan} and Convectively-Dominated flows
\citep{quataert_cdaf} were invented to explain this lower density. Magnetized solution without angular momentum does
well the same job. Let me consider the reference magnetized model with $\sigma_\infty=1,$ $\gamma=1,$
$\xi_\infty=0.025,$ $l=0,$ 1-T relativistic equation of state. The ratio of density in a reference magnetized model to
density in a non-magnetized solution is
\begin{equation}\label{densities_ratio}
\frac{\rho_{\rm magn}}{\rho_{\rm nonmagn}}\approx 0.27 \quad{\rm at}\quad 10r_{\rm g}.
\end{equation} Density in a magnetized model is much lower than in a non-magnetized one. However, all types of models can be made to
fit the data by adjusting temperature \citep{quataert_farad}, whether advection dominated or convection or outflow
dominated.

Faraday rotation of submillimeter radiation offers a good differentiation mechanism between ADAF flows and flows with
outflows or convection. Rotation measure is proportional to both magnetic field and electron density and has a
relativistic temperature factor \citep{marrone}. Model $B$ predicts magnetization $\sigma=0.7$ and number density
$n=2\cdot10^{7}\rm cm^{-3}$ at $3r_{\rm g}$ that is consistent with \citep{hawley}. The observed Faraday rotation measure is $\rm
RM=-6\cdot10^{-5}rad ~m^{-1}.$ \citep{marrone}. Fitting the relativistic rotation measure for temperature gives $T_e=4\cdot10^{10}$~K in excellent
agreement with \citep{sharma}. Accretion rate in the reference model is about $9\cdot10^{-7}M_\odot \rm year^{-1},$
what is $30$ times lower than in \citep{sharma}. However, the electron density in my model is close to that in the
rotating model \citep{sharma}, because inflow velocity in the rotating model is $\alpha$ times lower.
For densities to agree I need $\alpha\sim 0.03$ that is somewhat smaller than found in numerical simulations
$\alpha\gtrsim0.2$ \citep{hawley}. This means my solution overestimates density $n$ by about a factor of $5,$ what results in larger then observed IR flux \citep{eckart}.
 Effects of angular momentum transport, outflows
\citep{yuan_jet} or conduction \citep{johnson} must come into play to allow for successful fitting for both IR flux and Faraday rotation measure.

\section{CONCLUSIONS}\label{sec_conclusion}
Though many ways of dealing with inefficient accretion were
invented, my approach is substantially different from all previous
efforts. \textbf{I elaborated the model that}
\begin{itemize}
\item{has very few free parameters,} \item{self-consistently includes averaged turbulence, combining geometrical effects of
freezing-in amplification with dissipation,} \item{ties evolution of random magnetic field and random velocity field to numerical
simulations,} \item{connects outer externally supported turbulence to inner self-sustained turbulence,} \item{predicts the
accretion rates $\dot{M}$ and flow patterns for the flows with negligible angular momentum,}\item{gives the order of magnitude
estimate of $\dot{M}$ for large angular momentum flows.}
\end{itemize}
\textbf{The model predicts}
\begin{itemize}
\item{accretion rate $\dot{M}$ of magnetized fluid $0.2\div 0.7$ of Bondi rate $\dot{M}_B$ even for small outer magnetization
$\sigma_\infty$,} \item{subequipartition magnetic field in the outer part of the flow and superequipartition in the inner part,}
\item{several times lower density than in Bondi model near the central object, what with addition of other effects would explain the observations of Sgr A*,} \item{half an
order of magnitude effect of different equations of state on the accretion rate,} \item{unimportance of magnetic helicity
conservation,} \item{ineffectiveness of convection. Convection and diffusion should be accounted for together.}
\end{itemize}
\textbf{The next version of the model will include}
\begin{itemize}
\item{more anisotropic effects, in particular, magneto-rotational
instability,}\item{two-temperature equations of state,} \item{full
treatment of angular momentum transport,} \item{diffusion of
momentum, heat and magnetic field.}
\end{itemize}

\section{ACKNOWLEDGEMENTS}
The author is grateful to Ramesh Narayan for fruitful discussions.
The author thanks Pascal Demoulin for useful comments about
magnetic helicity and Ya. N. Istomin for general comments.

%\end{multicols}

\appendix

\section{ANALYTICAL TESTS}\label{section_analytic_tests} Let me consider my model in
anisotropic incompressible case of box turbulence.  I substitute $-v\partial/\partial r=d/dt$ in equations (\ref{B_r}),
(\ref{B_perp_r}), (\ref{u_r}) and set $r=\rm const.$ The box has infinite volume. I express some of unknown $c_{xx}$ in terms of
known $\hat{c}_{xx}$ from equations (\ref{hats}). The system now reads
\begin{mathletters}\label{incompressible_system}
\begin{equation}\label{B_r_box}
\frac{d(v_{Ar}^2)}{d t}=\frac{(\hat{c}_{Bu} v_{Ar}^2 + 2c_{Bu2}
(v_{A\perp}-v_{Ar})v_{Ar})u-(\sqrt{3}\hat{c}_{BB} v_{Ar} + 2
c_{BB2} (v_{A\perp}-v_{Ar}))v_{Ar}^2}{L},
\end{equation}
\begin{equation}\label{B_perp_box}
\frac{d(v_{A\perp}^2)}{d
t}=\frac{(\hat{c}_{Bu}v_{A\perp}^2+c_{Bu2}(v_{Ar}-v_{A\perp})v_{A\perp})u-(\sqrt{3}\hat{c}_{BB}v_{A\perp}
+ c_{BB2}(v_{Ar}-v_{A\perp}))v_{A\perp}^2}{L},
\end{equation}
\begin{equation}\label{u_box}
\frac{d(u^2)}{d
t}=\frac{(\hat{c}_{uB}(v_{Ar}^2+2v_{A\perp}^2)-c_{uB2}(v_{Ar}-v_{A\perp})^2)u
-c_{uu}u^3}{L}.
\end{equation}
\end{mathletters}
I need to determine three coefficients $c_{BB2},$ $c_{uB2},$ and
$c_{Bu2}$ and prove the entire system
(\ref{incompressible_system}) makes sense.

There are three kinds of analytical tests divided by the degree of their certainty. The tests from the first group have solid
physical grounds. The tests from the second group represent how turbulence is believed to work, these are the general relations
with clear physical insight. The third group of tests consists of the order of magnitude relations and the disputable ideas.

The tests of the first group are proven to work. Only one test of this kind can be applied to our system. This is the energy
decay test. Free incompressible MHD turbulence has decreasing with time total energy, because energy decrease corresponds to the
increase of entropy of the system gas/magnetic field \citep{landau8}.
\begin{equation}
\frac{d}{dt}\left(\frac{v_{Ar}^2+2v_{A\perp}^2+u^2}{2}\right)<0
\quad {\rm for~at~least~one~of} \quad v_{Ar}, v_{A\perp}, u \quad
{\rm non-zero.}
\end{equation}
I take sum with proper coefficients of the right-hand sides of
system (\ref{incompressible_system}). Then I maximize it with
respect to $v_{A\perp}/v_{Ar}$ and $v_{A}/u.$ I find that when
\begin{equation}\label{cond_energy}
2c_{Bu2}+c_{uB2}\geq-2.2,
\end{equation}
total energy decreases with time for any non-zero $v_{Ar},$ $v_{A\perp},$ and $u.$ Let me remind the reader that all these
velocity are non-negative according to definitions (\ref{Alfven_speed}). Condition (\ref{cond_energy}) is weak. Some tests from
the second and the third categories constrain $c_{uB2}$ and $c_{Bu2}$ better, thus making equation (\ref{cond_energy}) valid.

The typical test of the second category deals with dynamo amplification of anisotropic field. Dynamo action not only amplifies
magnetic field, but also isotropizes it. I take isotropization condition to be
\begin{equation}
\frac{d(v_{Ar}-v_{A\perp})}{dt(v_{Ar}-v_{A\perp})}\leqslant 0.
\end{equation} Taking expressions for derivatives from system (\ref{incompressible_system}) I arrive at
\begin{mathletters}\label{dynamo_test}
\begin{equation}\label{dt}
 (\hat{c}_{Bu}-3c_{Bu2})u -
\sqrt{3}c_{BB2}(v_{Ar}+v_{A\perp})+c_{BB2}(2v_{Ar}+v_{A\perp})\leqslant
0
\end{equation} This condition should hold when any speed in inequality (\ref{dt}) is much
larger then two others. Therefore, inequality (\ref{dt}) is
equivalent to
\begin{equation}\label{cond_dynamo}
\hat{c_{Bu}}<3c_{Bu2}, \qquad
c_{BB2}<\frac{\sqrt{3}}{2}\hat{c}_{BB}.
\end{equation}
\end{mathletters}

Another second category dynamo test states that magnetic field should always increase, if dynamo operates without dissipation or
any energy transfer. This occurs when Alfven speeds are much smaller than turbulent velocity field $u.$ Positive amplification
condition then reads
\begin{equation}\label{dynamo_test_2}
\frac{d v_{Ar}^2}{dt~v_{Ar}^2}>0, \qquad \frac{d v_{A\perp}^2}{dt~
v_{A\perp}^2}>0.
\end{equation} Taking the expressions for derivatives from system (\ref{incompressible_system}) and applying the
limit $v_{Ar}\ll u$ and $v_{A\perp}\ll u$ I obtain that
inequalities (\ref{dynamo_test_2}) are valid for any balance
between $v_{Ar}$ and $v_{A\perp}$ when
\begin{equation}\label{cond_dynamo_2}
\hat{c}_{Bu}>2c_{Bu2}.
\end{equation}
 Inequalities  (\ref{cond_dynamo}) and
(\ref{cond_dynamo_2}) give tight constrains on $c_{Bu2}.$

The similar test exists for the random velocity. Magnetic field is supposed to increase the turbulent velocity in the limit
$v_{Ar}\sim v_{A\perp}\gg u.$ The correspondent condition
\begin{equation}\label{velo_test}
\frac{d}{d t}\left(\frac{u^2}{2}\right)>0 \quad {\rm for} \qquad
v_{Ar}\sim v_{A\perp}\gg u
\end{equation} reduces for system (\ref{incompressible_system}) to the condition of constant positive
acceleration that initially steady magnetic field applies to
matter. Finally
\begin{equation}\label{cond_velo}
c_{uB2}<\hat{c}_{uB}.
\end{equation}

Decay of isotropic MHD turbulence offers the following test of the second kind. Numerical simulations show equality of magnetic
field dissipation rate and random velocity dissipation rate (\ref{c_BB}) when initial magnetic energy equals initial kinetic
energy. However, this equality should be stable, otherwise kinetic and magnetic energy would diverge from each other after any
perturbation and equality of $u$ and $v_{A}$ would not have been observed. Stability condition is
\begin{equation}\label{cond_stable}
\frac{d(v_{Ar}^2+2v_{A\perp}^2-u^2)}{dt~(v_{Ar}^2+2v_{A\perp}-u^2)}<0
\end{equation} for $v_{Ar}=v_{A\perp}=u.$

The are no more proven or justified assumptions I can make.  I need to make use of inequalities (\ref{cond_energy}),
(\ref{cond_dynamo}), (\ref{cond_dynamo_2}), (\ref{cond_velo}),  and (\ref{cond_stable}) and apply unjustified tests. I take the
value of $c_{Bu2}$ to be in the middle of the allowed interval
\begin{equation}\label{c_Bu2}
c_{Bu2}=\frac12\left(\frac12+\frac13\right)\hat{c}_{Bu}\approx0.29.
\end{equation} The value of $c_{uB}$ is small compared to the
values of other coefficients. There is no physical sense in the sharp increase of $u^2$ build-up when magnetic field becomes
anisotropic that would  be the case for $c_{uB2}\ll (-\hat{c}_{uB})$ Turbulent velocity may be expected to increase regardless of
the direction of magnetic field in equation (\ref{u_box}). This idea leads to $|c_{uB2}|<\hat{c}_{uB}.$ I take
\begin{equation}\label{c_uB2}
c_{uB2}=0
\end{equation} for the simple estimate. Similar estimate allows me
to set
\begin{equation}\label{c_BB2}
c_{BB2}=0.
\end{equation} In this case isotropization of magnetic field has a
timescale about the dissipation timescale.

\section{SELF-SIMILAR SOLUTION}\label{self_similar_solution} Let me
describe the self-similar solution, when the differential system of equations (\ref{system}) can be reduced to the algebraic
system. I set the proper scalings of quantities with radius and make weak additional assumptions. I introduce the standard
dimensionless variables $T(x),\rho(x), L(x), aa(x), bb(x), pp(x), vel(x)$ to replace, respectively, $T(r),\rho(r), L(r),$ $u(r),
v_{Ar}(r), v_{A\perp}(r), v(r)$ as follows:

\begin{alignat}{3}\label{dimensionless}
T(r)&=T_\infty T(x), &\qquad v(r)&=vel(x)\left(\frac{2R
T(x)}{\mu}\right)^{1/2},  &\qquad L(r)&=(r/x)L(x),
\\u(r)&=aa(x) \left(\frac{2R T(x)}{\mu}\right)^{1/2}, &\qquad
v_{Ar}(r)&=bb(x) \left(\frac{2R T(x)}{\mu}\right)^{1/2}, &\qquad
v_{A\perp}(r)&=pp(x) \left(\frac{2R T(x)}{\mu}\right)^{1/2}.
\nonumber
\end{alignat}
Radius is normalized to Bondi radius (eq. [\ref{bondi}]) as
$r=r_B~x$. The natural power-law radial dependencies of these
quantities (\ref{dimensionless})
\begin{alignat}{3}\label{scaling}
T(x)&=T_{\rm SS}x^{-1}, &\qquad vel(x)&=v_{\rm SS}x^{-1/2},
&\qquad L(x)&=\gamma, \\aa(x)&=u_{\rm SS}x^{-1/2}, &\qquad
bb(x)&=v_{Ar\rm SS}x^{-1/2}, &\qquad pp(x)&=v_{A\perp\rm
SS}x^{-1/2} \nonumber
\end{alignat}
make my system of equations (\ref{system}) independent of $x$ under the following restrictions:
\begin{itemize}\label{restrictions}
\item{gravity is Newtonian,} \item{external turbulence driving is
negligible,} \item{equation of state is non-relativistic.}
\end{itemize}
These assumptions are valid in the intermediate region $10^3 r_{\rm g}\lesssim r \lesssim 0.1 r_B$. Gravity is Newtonian for
$r\gg r_{\rm g}.$ Turbulence driving is mainly internal for $r\lesssim 0.1r_B$ (see subsections (\ref{maximum_rate_solution}),
(\ref{physical_solution}) and Fig.~\ref{fig2}b, Fig.~\ref{fig5}b). Electrons become relativistic at around $10^3r_{\rm g}.$ The
found range of $r$ where all above assumptions hold is small. I can instead consider a non-relativistic equation of state with
$w=w_{NR}$ (eq. [\ref{advect1}]) everywhere. This makes standard self-similar solution possible from $0.1r_B$ down to several
Schwarzschild radii $r_{\rm g}.$

Dimensionless magnetic helicity $\xi$ appears to be constant in self-similar regime. Relations (\ref{L_r1}), (\ref{hel_r1}), and
(\ref{magnetization}) lead to
\begin{equation}\label{ss_helicity}
\xi=\frac{3\sigma_\infty}{4 T_{SS}(v_{Ar SS}^2+2v_{A\perp SS}^2)}\xi_\infty.\end{equation} Continuity equation (\ref{mdot_r1})
can be used to obtain the scaling of density $\rho\sim x^{-3/2}.$ Heat balance equation (\ref{entropy}) reduces to the equality
of radial and total perpendicular magnetic fields
\begin{equation}\label{ss_entropy}v_{A SS}^2=2v_{A\perp
SS}^2.\end{equation} Euler equation (\ref{Euler1}) gives the formula for self-similar temperature
\begin{equation}\label{ss_temperature}T_{SS}=5/(15+10u_{SS}^2+9v_{Ar SS}^2+6v_{A\perp
SS}^2+6v_{SS}^2).\end{equation} Turbulence evolution equations
(\ref{B_r}), (\ref{B_perp_r}), (\ref{u_r}) are now treated without
source terms. They give, correspondingly, three relations
\begin{eqnarray}\label{ss_turbulence}
2 u_{SS} v_{ArSS} c_{Bu11} -
 2 v_{ArSS}^2 c_{BB11} \exp(-\xi) + 4 u_{SS} c_{Bu22} v_{A\perp SS} + 3 v_{ArSS} v_{SS}
  \gamma=0, \nonumber\\
 2 u_{SS} (v_{Ar SS} c_{Bu22} + (c_{Bu11} +
c_{Bu22}) v_{A\perp SS}) -
 v_{A\perp SS} (2 c_{BB11} \exp(-\xi) v_{A\perp SS} + 3 v_{SS}
 \gamma)=0, \\
- u_{SS}^2 c_{uu} +  c_{uB11} \exp(-\xi)(v_{ArSS}^2 + 2 v_{A\perp
SS}^2)=0,\nonumber
\end{eqnarray} where definitions of Alfven and turbulent velocities (\ref{Alfven_speed}) are used.

Let me first set magnetic helicity to zero $\xi=0$ and consider four equations (\ref{ss_entropy}) and (\ref{ss_turbulence}) on
four velocities $v_{SS},$ $u_{SS},$ $v_{Ar SS},$ $v_{A\perp SS}.$ The only solution of this system has all the velocities
identical zeroes. No self-similar solution is possible for zero magnetic helicity $\xi$.

However, the non-linear algebraic system of equations on $\xi$ and velocities (\ref{ss_helicity}), (\ref{ss_entropy}),
(\ref{ss_turbulence}) possesses a non-trivial self-similar solution. For the full system (\ref{system}) I need the additional
condition to determine the accretion rate and solve for radial dependencies of quantities. This condition is either condition for
maximum accretion rate (\ref{sonic}) or condition for effective angular momentum transport (\ref{condition}). I can transform
both into self-similar form.
\begin{mathletters}\label{ss_cond}
Maximum $\dot{M}$ condition (\ref{sonic}) reads
\begin{equation}\label{ss_sonic}
5+10u_{SS}^2+12v_{A\perp SS}=6v_{SS}^2.
\end{equation}
Effective angular momentum transport condition (\ref{condition}) gives
\begin{equation}\label{ss_condition}
\frac{\sqrt{5/3}~v_{SS}}{4v_{Ar SS}~v_{A\perp
SS}\sqrt{T_{SS}}}\le1
\end{equation}
regardless of circularization radius $r_{\rm cir}.$
\end{mathletters}

Let me first find the self-similar solution in case of large angular momentum. I solve equality in relation (\ref{ss_condition})
and 5 equations (\ref{ss_helicity}), (\ref{ss_entropy}), (\ref{ss_temperature}), (\ref{ss_turbulence}) for 7 quantities $\xi,$
$T_{SS},$ $u_{SS},$ $v_{Ar SS},$ $v_{A\perp SS},$ $\gamma~v_{SS},$ and the product $\sigma_\infty\xi_\infty.$ I normalize the
results to free-fall velocity (eq. [\ref{v_ff}]) to be able to directly compare with the numbers on Figure~\ref{fig5}b:
\begin{alignat}{3}\label{ss_results}
\frac{c_s(r)}{v_{ff}(r)}&=0.58, \quad
\frac{u(r)}{\sqrt{3}v_{ff}(r)}&=0.0094, \quad
\frac{v_{Ar}(r)}{v_{ff}(r)}&=0.041, \\
\frac{v_{A\perp}(r)}{v_{ff}(r)}&=0.029, \quad
\frac{v(r)}{v_{ff}(r)}&=0.0033, \quad
\sigma_\infty\xi_\infty&=0.00718
\end{alignat} for $r\gg r_{\rm g}.$ Figure~\ref{fig5}b shows profiles of
velocities for the reference model with $\sigma_\infty=1,$ $\xi_\infty=0.025,$ $\gamma=1.$ The actual velocities on the inner
boundary at $r=3\times10^{-4}r_B=90r_{\rm g}$ are
\begin{alignat}{3}\label{ss_actual}
\frac{c_s(r)}{v_{ff}(r)}&=0.58, \quad
\frac{u(r)}{\sqrt{3}v_{ff}(r)}=0.0033, \quad
\frac{v_{Ar}(r)}{v_{ff}(r)}&=0.076, \\
\frac{v_{A\perp}(r)}{v_{ff}(r)}&=0.024, \quad
\frac{v(r)}{v_{ff}(r)}=0.0051. \quad \nonumber
\end{alignat}
The reference model has $\sigma_\infty\xi_\infty=0.025$ about $3$ times larger than in self-similar solution (\ref{ss_results}),
magnetic field in the reference  model is stronger. Therefore, higher values of all characteristic velocities are expected in the
actual solution (\ref{ss_actual}). I obtain inflow velocity $v$ and radial Alfven speed $v_{Ar}$ correspondingly $1.5$ and $1.8$
times higher for solution (\ref{ss_actual}). Sonic speeds are the same in self-similar (\ref{ss_results}) and actual
(\ref{ss_actual}) solutions, because almost all gravitational energy goes into thermal energy in both cases. However,
perpendicular Alfven velocity $v_{A\perp}$ and turbulent velocity $u$ do not qualitatively agree with self-similar solution. They
are correspondingly $1.2$ and $2.8$ times lower in the actual solution (\ref{ss_actual}). The naive estimate for accretion rate
is
\begin{equation}\label{ss_mdot}
4\pi\rho_\infty v(r_B) r_B^2\approx 0.05\dot{M}_B.
\end{equation} This appears to be $8$ times larger than the actual
accretion rate $0.0061\dot{M}_B.$ Velocity near Bondi radius (eq. [\ref{bondi}]) is much smaller than self-similar value, what
leads to an overestimate of $\dot{M}.$ Thus, self-similar solution can give an order of magnitude estimates for all
characteristic velocities of the flow and even for accretion rate $\dot{M}.$ However, self-similar solution has only $2$ free
parameters instead of $3$, because $\sigma_\infty\xi_\infty$ is treated as one constant. Therefore, solution of the full system
(\ref{system}) is required to probe the entire parameter space and to achieve more precise results.

Self-similar solution in case of maximum rate flow with condition (\ref{ss_sonic}) does not exist. The formal solution of
equations (\ref{ss_sonic}), (\ref{ss_helicity}), (\ref{ss_entropy}), (\ref{ss_turbulence}) leads to negative product
$\sigma_\infty\xi_\infty.$ The absence of self-similar solution in this case is reasonable, since the actual solution does not
exhibit self-similar scalings (Fig.~\ref{fig2}b).

\section{CONVECTION}\label{section_convection}
Let me elaborate the stability criterion against convection in my model. As I noted in the main text
(subsection~\ref{subsection_convection}), small scale perturbations of quantities are smeared out by diffusion. Thus
high-frequency analysis by \citet{scharlemann} is not appropriate to determine the convective stability. Timescale of diffusion
$\tau_{\rm diff}$ is
\begin{equation}\label{time_diff}
\tau_{\rm diff}\sim \frac{h}{u},
\end{equation} where $l$ is the scale of perturbation. As $h$
decreases, diffusion time also decreases and becomes smaller than perturbation growth timescale $\tau_{\rm grow}.$ If $\tau_{\rm
diff}< \tau_{\rm grow},$ convection is ineffective that is likely to happen at small scales $h.$ Thus I need to consider the
motion of the large blobs of the size $h\sim L.$

I consider a blob of plasma displaced at some small $\Delta r$ from its equilibrium position (Fig.~\ref{fig7}). The density of
the blob itself changes by $\Delta \rho_{\rm blob},$ when it is moved. The density of outer medium changes by $\Delta \rho_{\rm
fluid}$ between two positions of the blob. The goal is to calculate the difference in density differences $\Delta \rho_{\rm
fluid}-\Delta\rho_{\rm blob}$ between the outer medium and the blob. Positive difference $\Delta \rho_{\rm fluid}-\Delta\rho_{\rm
blob}>0$ for positive $\Delta r>0$ implies convective instability. Rising blob of gas is rarified compared to the fluid and
buoyant. The results for $\Delta \rho$ may be affected by external driving that is somewhat artificial in my model. Thus I need
to calculate $\Delta \rho$ in the inner accretion region where external driving is not important. Motion of the blob is adiabatic
and governed by the same adiabatic dynamical equations (\ref{Euler1}) and (\ref{advect1}), as the rest of the fluid. I neglect
energy, associated with gas regular velocity $v$. Term $v^2$ cannot be neglected only in the region, where $v$ approaches sound
speed $c_s.$ However, convection ceases if $v\sim c_s$ \citep{narayan_CDAF}. I denote by index $A$ physical quantities in the
blob and by index $F$ quantities in the rest of the fluid.

Euler equation (\ref{Euler1}) results in the following equations on differences in the blob
\begin{mathletters}\label{radial_pressure_diff}
\begin{equation}
\frac{R}{\mu}\Delta_A (\rho T)+\frac13 \Delta_A(\rho
u^2)+\frac{1}{r^2}\Delta_A(r^2 \rho
v_{A\perp}^2)-\frac{1}{2r^4}\Delta_A(r^4\rho v_{A}^2)=0
\end{equation}
and in the fluid
\begin{equation}
\frac{R}{\mu}\Delta_F (\rho T)+\frac13 \Delta_F(\rho
u^2)+\frac{1}{r^2}\Delta_F(r^2\rho v_{A\perp}^2)-\frac{1}{2r^4}
\Delta_F(r^4\rho v_{A}^2)=0.
\end{equation}
In both equations I take variations between quantities at $r+ \Delta r$ and $r.$
\end{mathletters}
I introduce the difference operator
\begin{equation}\label{diff_operator}
\Delta()=\Delta_F()-\Delta_A()
\end{equation}
 and calculate the variations of
all quantities between the fluid and the blob. Subtracting equation (\ref{radial_pressure_diff}b) from equation
(\ref{radial_pressure_diff}a), I find the radial pressure balance in the first order in $\Delta r$
\begin{equation}\label{radial_balance}
\frac{R}{\mu}\Delta(\rho T)+\frac13 \Delta(\rho u^2)+\Delta(\rho
v_{A\perp}^2)-\frac12 \Delta(\rho v_{A}^2)=0.
\end{equation}

Blob of plasma should be in equilibrium also in perpendicular direction, not only in radial direction. I use the same technique
to deduce it, as I used to derive the radial force equation (\ref{Euler}) from general momentum equation (\ref{gen_momentum}).
Component $\theta$ of magnetic force in equations (\ref{gen_Euler}) and (\ref{gen_momentum}) reads $F_\theta=[{\bf
B\times[\nabla\times B]}]_\theta/(4\pi\rho).$ I subtract $B_\theta({\bf  \nabla\times B})/(4\pi\rho)$ from it and average over
$\phi$ direction. I obtain
\begin{equation}\label{theta_force}
F_\theta=\frac{(B_r^2)'_\theta}{8\pi\rho r}
\end{equation} for $B_\theta^2=B_\phi^2$ and $B_rB_\theta=0$ on
average over $\phi.$ The final form of force balance in $\theta$ direction is
\begin{equation}\label{perp_pressure_diff}
\frac{\partial}{\partial \theta}\left(\frac{R}{\mu}\rho T+\frac13
\rho u^2 + \frac12 \rho v_{A}^2\right)=0.
\end{equation} Perpendicular force balance (\ref{perp_pressure_diff}) has the same form in any direction perpendicular to the radial vector
owing to the symmetry of the problem. I apply operator $\Delta$ (eq. [\ref{diff_operator}]) to the integral form of perpendicular
pressure balance and get
\begin{equation}\label{perp_balance}
\frac{R}{\mu}\Delta(\rho T)+\frac13 \Delta(\rho u^2) + \frac12
\Delta(\rho v_{A}^2)=0.
\end{equation}
Heat balance equation (\ref{entropy}) gives the third relation
\begin{equation}\label{entropy_balance}
\frac{R}{\mu}\left(\frac32\Delta T -\frac{\Delta\rho}{\rho}
T\right)+ \left(u\Delta u -
\frac{u^2}{3}\frac{\Delta\rho}{\rho}\right) +
\rho\Delta\left(\frac{v_{A\perp}^2}{\rho}\right)+\frac{1}{2\rho}
\Delta(\rho v_A^2)=0.
\end{equation}

 Expansion or contraction of a blob is non-uniform.
Perpendicular $b$ and parallel $a$ sizes (Fig.~\ref{fig7}) deform in different ways. Continuity equation for the fluid
(\ref{mdot_r}) can be written as
\begin{mathletters}
\begin{equation}\label{medium_cont}
\frac{\Delta_F\rho}{\rho}+\frac{\Delta_F v}{v}+2\frac{\Delta
r}{r}=0
\end{equation}
I consider the parcel with constant mass $m=\rho V.$ Therefore
\begin{equation}\label{blob_cont}
\frac{\Delta_A\rho}{\rho}+\frac{\Delta a}{a}+2\frac{\Delta b}{b}=0
\end{equation}
is the continuity relation for the parcel.
\end{mathletters}
Finally I subtract equation (\ref{medium_cont}) from equation
\ref{blob_cont} and obtain
\begin{equation}\label{cont_balance}
\frac{\Delta\rho}{\rho}+\frac{\Delta v}{v}+2\frac{\Delta
r}{r}-\frac{\Delta a}{a}-2\frac{\Delta b}{b}=0
\end{equation} for the change of
density according to definition (\ref{diff_operator}). Inflow velocity $v$ is clearly associated with the fluid, but I omit
subscript $F$ at $v$. I also omit subscript $A$ at dimensions of the blob.

Now I need to quantify the variation of the turbulent magnetic field and the random velocity. I assume that the blob moves at a
speed $V(r)$ much higher than the inflow velocity $V(r)\gg v(r),$ therefore magnetic field does not dissipate in the parcel.
Differences of turbulence evolution equations (\ref{B_r1}), (\ref{B_perp_r1}), and (\ref{u_r1}) are
\begin{mathletters}\label{turbulence_balance}
\begin{equation}
2u \Delta u - \frac23 u^2 \frac{\Delta \rho}{\rho} = \frac{\Delta
r}{v L} (c_{uu}u^3-c_{uB11}(v_{A}^2+2v_{A\perp}^2)u \exp(-\xi))
\end{equation}
\begin{equation}
 \Delta(\rho v_{A}^2)
+ 4\rho v_{A}^2\left(\frac{\Delta r}{r}-\frac{\Delta
b}{b}\right)=\frac{\rho \Delta r}{v
L}(c_{BB11}v_A^3\exp(-\xi)-(c_{Bu11}v_{Ar}^2+2c_{Bu22}v_{Ar}v_{A\perp})u)
\end{equation}
\begin{eqnarray}
&\Delta(\rho v_{A\perp}^2) + 2\rho v_{A\perp}^2\left(\frac{\Delta
r}{r}+\frac{\Delta v}{v}-\frac{\Delta a}{a}-\frac{\Delta
b}{b}\right)=\\&= \frac{\rho \Delta r}{v
L}(c_{BB11}v_{A\perp}^3\exp(-\xi)-((c_{Bu11}+c_{Bu22})v_{A\perp}^2-c_{Bu22}v_{Ar}v_{A\perp})u).\nonumber
\end{eqnarray}
\end{mathletters}

 Magnetic helicity variation does not directly influence the dynamics of the blob. Solving the system of 7
equations (\ref{radial_balance}), (\ref{perp_balance}),
(\ref{entropy_balance}), (\ref{cont_balance}),
(\ref{turbulence_balance}abc) on 7 quantities $\Delta T, \Delta
\rho, \Delta v_A, \Delta v_{A\perp},$ $\Delta u, \Delta a, \Delta
b,$ I obtain
\begin{equation}\label{correct_rho}
\frac{\Delta{\rho}_{\rm correct}}{\rho \Delta r}\approx
v_{Ar}\frac{2.02\exp(-\xi)v_{Ar}v_{A\perp}(v_{Ar}+2v_{A\perp})-
u(0.39(v_{Ar}^2v_{A\perp}+v_{A\perp}^3)+v_{Ar}(1.21v_{A\perp}^2-0.63u^2))}{c_s^2
L v(v_{Ar}^2+v_{A\perp}^2)}.
\end{equation}The actual expression is much longer. I take only the largest
terms in the numerator and the denominator.

Let me compare this result (eq. [\ref{correct_rho}]) with the naive estimate, when magnetic field dissipation increases gas
internal energy only \citep{bisnovat_74}, and gas pressure balance is used instead of parallel and perpendicular pressure
balances (\ref{radial_balance}), (\ref{perp_balance}). Gas pressure balance is
\begin{equation}\label{naive_pressure}
\Delta(\rho T)=0.
\end{equation}
Naive heat balance (\ref{advect}) for the unit mass is
\begin{equation}\label{naive_heat}
\frac{R}{\rho\mu}\left(\frac32 \rho\Delta T-T\Delta\rho
\right)\approx\frac{\Delta r}{L v}(0.41v_{Ar}^2u+1.16v_{Ar}u
v_{A\perp}+1.4u
v_{A\perp}^2-3.03(v_{Ar}^3+2v_{A\perp}^3)\exp(-\xi)-1.14u^3)
\end{equation}
Eliminating $\Delta T$ from relations (\ref{naive_pressure}) and
(\ref{naive_heat}), I find
\begin{equation}\label{naive_rho}
\frac{1}{\rho}\frac{\Delta{\rho}_{\rm naive}}{\Delta
r}\approx\frac{
0.61(v_{Ar}^3+2v_{A\perp}^3)\exp(-\xi)+0.23u^3-0.82v_{Ar}^2u-0.23v_{Ar}u
v_{A\perp}-0.28u v_{A\perp}^2}{c_s^2 L v}
\end{equation}

\begin{figure}[h]
\epsscale{.85} \plotone{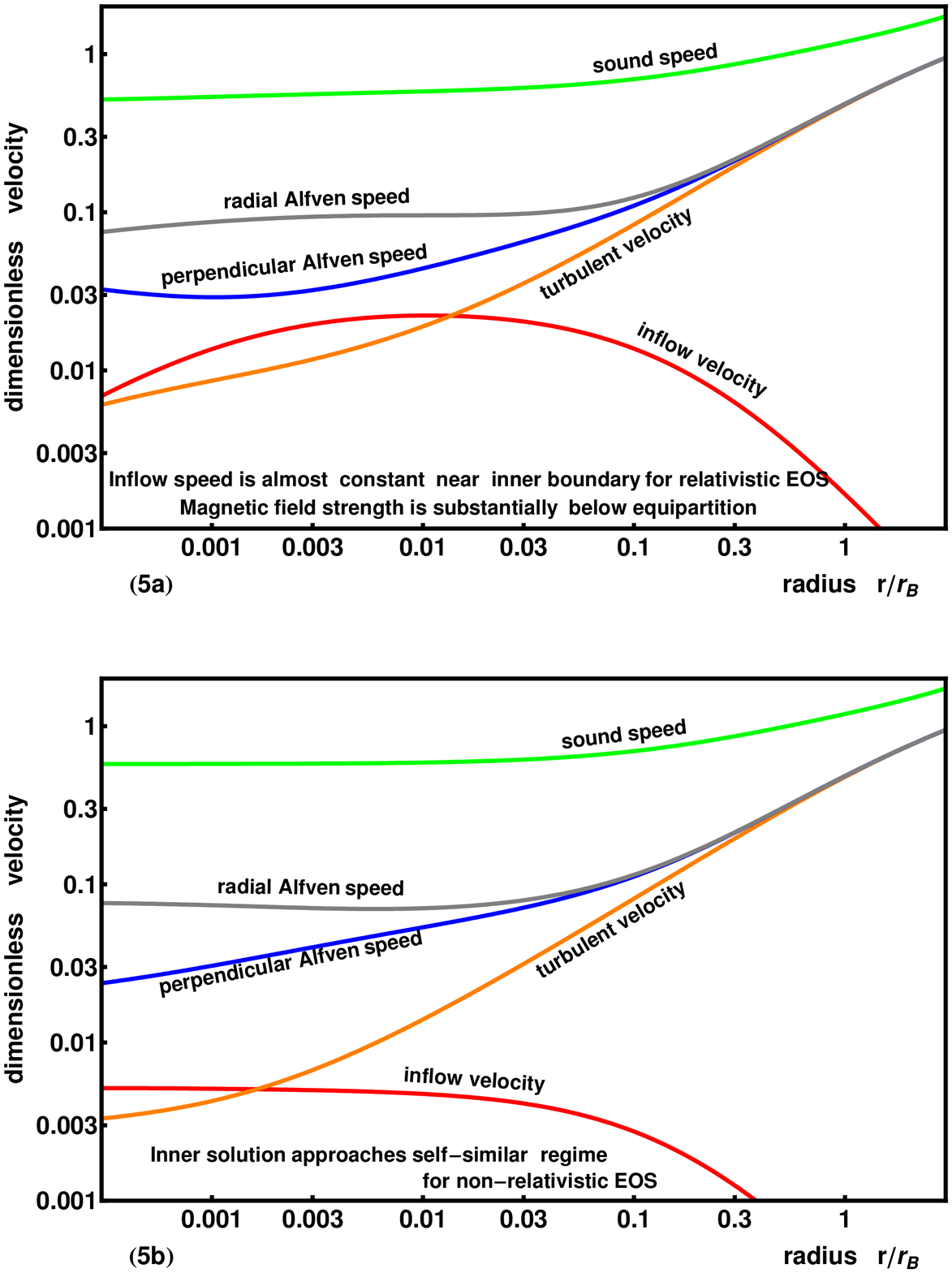}
 \caption{Flow velocities, normalized to free-fall speed verses radius for \textbf{solution with angular momentum
 transport}: sound speed, inflow velocity, radial Alfven speed, 1-D perpendicular Alfven speed, turbulent velocity. Parameters $\sigma_\infty=1$,
$\gamma=1$, $\xi=0.025$, $r_{\rm cir}=10^3r_{\rm g }$. Relativistic 1-T EOS on Figure~\ref{fig5}a, non-relativistic 1-T EOS on
Figure~\ref{fig5}b.}
 \label{fig5}
\end{figure}

\begin{figure}[h]
\plotone{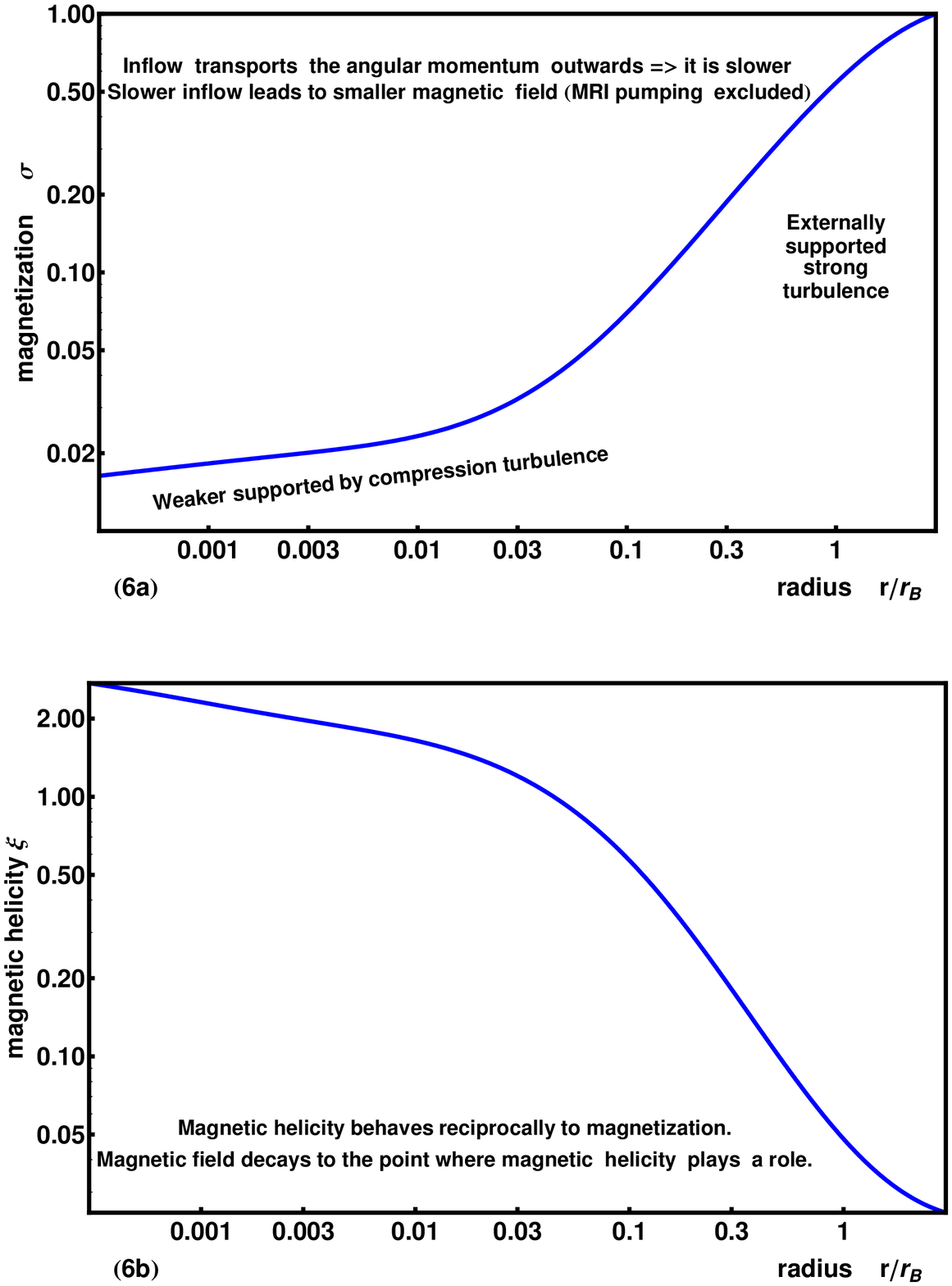} \epsscale{1}
 \caption{Magnetization $\sigma$ verses dimensionless distance from the compact object $r/r_B$ is on Figure~\ref{figang}a.
Dimensionless magnetic helicity $\xi$ verses dimensionless distance from the compact object $r/r_B$ is on Figure~\ref{figang}b.
Both are for \textbf{solution with angular momentum transport}. Circularization radius is $r_{\rm cir}=10^3r_{\rm g }.$}
 \label{figang}
\end{figure}

\begin{figure}[h]
\plotone{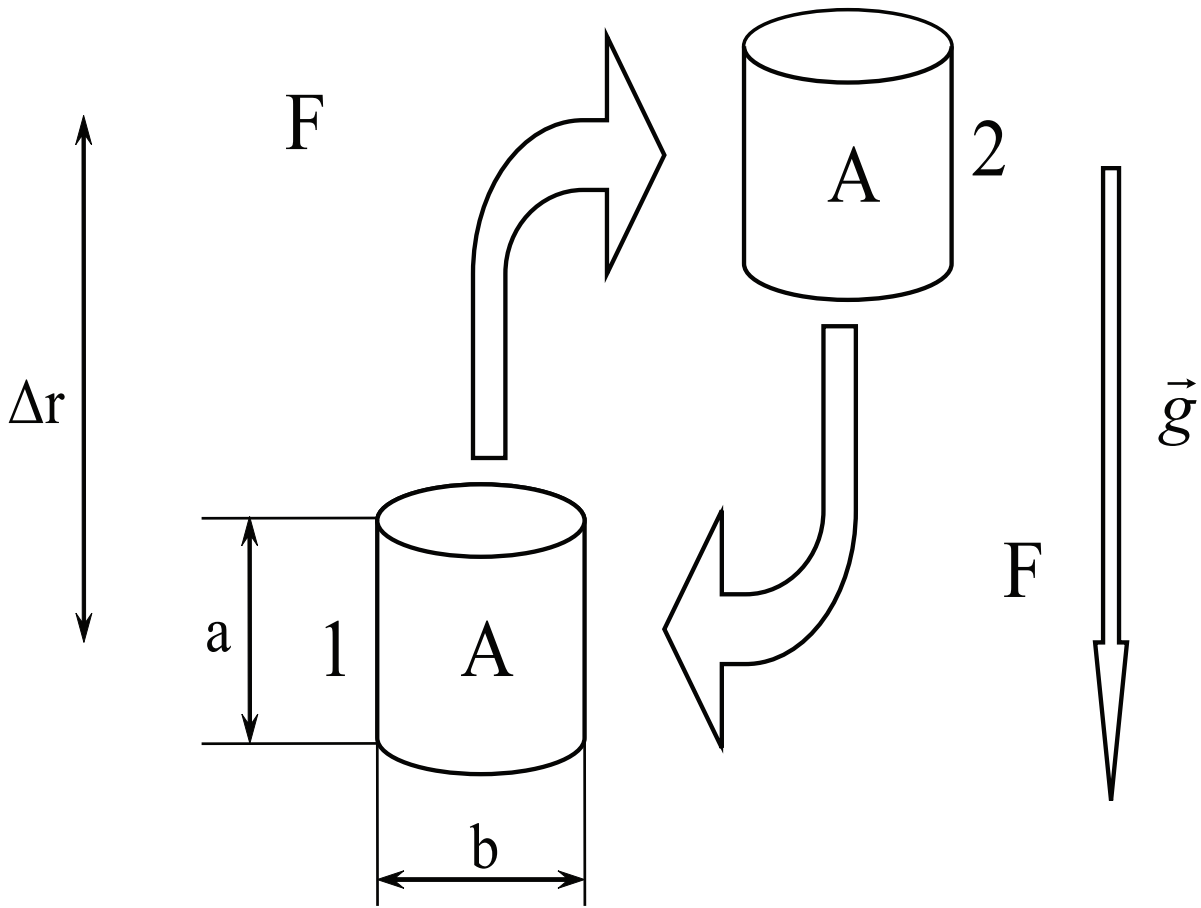}
 \caption{Scheme of convection. Large magnetized blob is in perpendicular and radial pressure balance. Energy does not dissipate inside the blob.}
 \label{fig7}
\end{figure}

I evaluate the convective derivatives of density (\ref{correct_rho}) and (\ref{naive_rho}) in the inner region of the reference
solution with angular momentum transport (subsection~\ref{physical_solution}). Parameters of the reference model are
$\xi_\infty=0.025$, $\sigma_\infty=1$, $\gamma=1,$ non-relativistic EOS. Correspondent velocities are shown on
Figure~\ref{fig5}b. I take the values (\ref{ss_actual}) of velocities and magnetic helicity on the inner boundary of integration
at $r=3\times10^{-4}r_B\approx90r_{\rm g}.$ Change of density appears to be negative $\Delta \rho<0$ for $\Delta r>0$ in the
result of full calculation (eq. [\ref{correct_rho}]). Naive calculation shows positive $\Delta \rho>0$ for $\Delta r>0.$
\begin{equation}\label{result_density}
\frac{\Delta\rho_{\rm correct}}{\Delta \rho_{\rm naive}}\approx -
0.2.
\end{equation} Naive calculation suggests that the flow is convectively unstable, whereas the full calculation under
reasonable assumptions indicates a convectively stable flow.

The calculated result (\ref{result_density}) is applicable only to the inner regions of solution with angular momentum transport
(subsection~\ref{physical_solution}). Excluded external driving is important in the outer regions. In turn, solution with maximum
accretion rate has large inflow velocity $v$ that approaches gas sound speed $c_s,$ and convection is suppressed (subsection~\ref{subsection_convection}).
 As a bottom line, either flow appears to be convectively stable on average or convection is suppressed in all calculated
solutions without electron conductivity. 

 However, numerical simulations by \citep{igumen06} of non-rotating flows find evidence of convection. This convection may be physical. My model averages heat from all
dissipation events over the fluid. Local reconnection events can lead to burst-type local heating that leads to buoyancy of
blobs. Also, magnetic buoyancy and diffusion play important role in transfer processes \citep{igumen06}. The correct inclusion of
convection, magnetic buoyancy and diffusion is the subject of future studies.

\clearpage

\end{document}